\documentclass[journal]{IEEEtran}
\usepackage{url} 
\usepackage{lineno}
\usepackage{soul}
\usepackage{graphicx}
\usepackage{subfig}
\usepackage{subfloat}
\usepackage{amsmath}
\usepackage{booktabs}
\usepackage{autobreak}
\usepackage{multicol}
\usepackage{multirow}
\usepackage{amsmath}
\usepackage{breqn}
\usepackage[colorlinks=true, allcolors=blue]{hyperref}
\usepackage[ruled,vlined]{algorithm2e}
\allowdisplaybreaks

\ifCLASSINFOpdf
\else
\fi

\begin{document}
\title{Physics-informed waveform inversion using pretrained wavefield neural operators}
\author{{Xinquan Huang, Fu Wang, Tariq Alkhalifah}
\thanks{Xinquan Huang is with the Physical Science and Engineering division, King Abdullah University of Science and Technology, and also with the School of Engineering and Applied Science, University of Pennsylvania, email: xinquan.huang@kaust.edu.sa,\\

Fu Wang and Tariq Alkhalifah are with the Physical Science and Engineering division, King Abdullah University of Science and Technology, email: fu.wang@kaust.edu.sa, tariq.alkhalifah@kaust.edu.sa.\\

© 2025 IEEE. Personal use of this material is permitted. Permission from IEEE must be obtained for all other uses, in any current or future media, including reprinting/republishing this material for advertising or promotional purposes, creating new collective works, for resale or redistribution to servers or lists, or reuse of any copyrighted component of this work in other works.
The published version is available at: 
\href{https://doi.org/10.1109/TGRS.2025.3624025}{https://doi.org/10.1109/TGRS.2025.3624025}}}%
\maketitle

\begin{abstract}
Full waveform inversion (FWI) is crucial for reconstructing high-resolution subsurface models, but it is often hindered, considering the limited data, by its null space resulting in low-resolution models, and more importantly, by its computational cost, especially if needed for real-time applications. 
Recent attempts to accelerate FWI using learned wavefield neural operators have shown promise in efficiency and differentiability, but typically suffer from noisy and unstable inversion performance. 
To address these limitations, we introduce a novel physics-informed FWI framework to enhance the inversion in accuracy while maintaining the efficiency of neural operator-based FWI. 
Instead of relying only on the L2 norm objective function via automatic differentiation, resulting in noisy model reconstruction, we integrate a physics constraint term in the loss function of FWI, improving the quality of the inverted velocity models. 
Specifically, starting with an initial model to simulate wavefields and then evaluating the loss over how much the resulting wavefield obeys the physical laws (wave equation) and matches the recorded data, we achieve a reduction in noise and artifacts. 
Numerical experiments using the OpenFWI and Overthrust models demonstrate our method's superior performance, offering cleaner and more accurate subsurface velocity than vanilla approaches. Considering the efficiency of the approach compared to FWI, this advancement represents a significant step forward in the practical application of FWI for real-time subsurface monitoring. 
  
\end{abstract}


\IEEEpeerreviewmaketitle

\section{Introduction}
Full waveform inversion (FWI) \cite{Tarantola1984}, in which we aim to iteratively converge to a model that matches the resulting simulated seismograms with the observed ones, can provide accurate and high-resolution subsurface parameter estimation, yielding insights into the understanding of Earth's interior. 
With the recent advancements in acquisition techniques, providing data support for FWI, and our constantly advancing computing capabilities provided by the likes of the graphics processing units (GPUs) \cite{Yang2015}, FWI has become increasingly accessible and has been used in exploration and global geophysics \cite{Sears2010,Oh2018,Borisov2020,liu_multiparameter_2021,thrastarson_reveal_2024}. 

However, in spite of improvements in acquisition, our data are still limited in spatial extent and bandwidth, often resulting in low-resolution models, especially at depth \cite{alkhalifah_full_2018}. This problem is often addressed with regularization, such as Tikhonov regularization \cite{golub_tikhonov_1999}, total variation regularization \cite{rudin1992nonlinear,chambolle2004algorithm}, and learned priors as regularization \cite{zhang_multiparameter_2018,wang_prior_2023}. These regularizations generally mitigate the model null space weakness, but they are as accurate or good as the assumption involved in them.
In addition, as the volume of acquired data grows further, particularly for time-lapse monitoring, FWI faces a dilemma in computational cost and efficiency, hampered by the time required to converge to an appropriate model. 
Delving into the framework of FWI, the computational burden is mainly caused by the required numerical simulations needed for forward modeling and for computing the adjoint. 
Reducing the computational cost of FWI has been a hot topic for the last few decades.
Marfurt (1984) \cite{Marfurt1984} proposed frequency-domain modeling, offering computational efficiency at the expense of sacrificing memory efficiency. 
There are plenty of other work aimed at reducing the number of simulations required in FWI such as source encoding \cite{Romero2000,Krebs2009}, stochastic shots selection, and compressive sensing \cite{Li2012,Zhu2015a}, or in utilizing other ways to efficiently represent the wavefield like the excitation approximation \cite{Chang1986,Kalita2016}, yielding less computational cost.
Although such developments offer improvements in the FWI computational cost, the computational burden remains prohibitive for any attempt on obtaining real-time results, such as those beneficial to time-lapse applications, as well as those used for monitoring.
 
Beyond traditional numerical methods, modern deep learning has emerged as a revolutionary framework to accelerate processes, including FWI.
These methods can be broadly categorized into two types: (1) direct data-driven approaches that learn an end-to-end inverse mapping from seismic data to model parameters, and (2) surrogate forward modeling approaches that accelerate iterative inversion by replacing numerical solvers with neural approximators.
Many attempts have been employed for direct data-driven inversion to instantly estimate the subsurface parameters \cite{yang_deep-learning_2019,li_deep-learning_2020,wu_inversionnet_2020}. 
In spite of their direct nature, those methods are not flexible or not easily generalizable during the application due to the need for an extensive mapping between the data space and the model space.
As neural networks are universal function/operator approximators \cite{Hornik1989,chen1995universal}, another routine is to utilize neural networks to surrogate the solving for partial differential equations (PDEs) to provide efficient modeling for those repeated simulations, a.k.a. neural PDE solvers, yielding efficient FWI.
Physics-Informed Neural Networks (PINNs), as one class of the neural PDE solvers for this purpose, have demonstrated promise in solving complex wave equations with meshless solutions \cite{Alkhalifah2021Wavefield,huang_pinnup_2022}. 
Despite their ability to solve frequency-domain wave equations, as they are currently stable and accurate for even multi-frequencies and multi-sources \cite{huang2022single}, they pose an inherent limitation for waveform inversion applications as these solutions correspond to a specific velocity model. 
Such a function learning method for only one instance of velocity requires plenty of time to retrain the model for each new velocity model. 
Efforts to address these limitations through efficient PINNs training with hash encoding \cite{huang_efficient_2024} and transfer learning \cite{Waheed2021} have shown promise but remain insufficient for time-lapse applications. 
Taufik et al. (2024) \cite{taufik_multiple_2024} proposed to incorporate the latent representation of velocities into the PINNs training, resulting in instant wavefield prediction for various velocities, but often limited to the relatively low-frequency components. 
Hence, PINN-based inversion, which jointly optimizes a neural network for the velocity model and another for wavefield representation, remains computationally slow and often produces overly smooth inversion results \cite{Song2022waveinv,rasht2022physics}.

Another type of neural PDE solvers, the neural operator \cite{li_fourier_2020}, though without the feature of meshless functional solutions of PINNs, has shown promising results in solving parametric PDEs, which is potentially ideal for real-time applications.
Wei and Fu (2022) \cite{wei2022small} developed a small-data-driven, physics-informed Fourier neural operator (FNO) framework for fast time-domain seismic simulations in complex media.
Building on this direction, Yang et al. (2023) \cite{yang_rapid_2023} proposed the FNO-based solver for time-domain wave equation modeling, and it has been applied to 3D seismic modeling \cite{lehmann2023fourier}.
Other studies \cite{zhang_learning_2023,li_solving_2023,zou_deep_2023} have employed FNO for frequency domain seismic modeling, as well as electromagnetic data simulation \cite{peng2022rapid}.
More recently, Wang et al. (2024) \cite{wang2024transfer} improved the accuracy and efficiency of FNO modeling by using transfer learning.
Those methods typically deal with the source locations and frequencies as binary masks and constant channels and feed them into the FNO for wavefield prediction, yielding suboptimal representation for the frequency domain wavefield.
Huang and Alkhalifah (2024) \cite{huang_learned_2024} proposed learned frequency-domain scattered wavefield modeling using FNO for various velocities, in which the frequencies and source locations are embedded in the input background wavefield, yielding good performance for multi-frequencies, multi-sources modeling for various velocities within the trained distribution. 
Then, we use these trained neural operators to provide efficient wavefield solutions for FWI \cite{yang_rapid_2023,zou_deep_2023}  and update the velocity model by back-propagating the residuals between the predictions from neural networks and the observed data through the neural operator using automatic differentiation \cite{baydin_automatic_2018}. 
Such a neural operator-based FWI framework is easily applied and flexible to incorporate other regularization terms. They also offer an opportunity to inject prior information into the velocity model through the training of the neural operator on a viable distribution of velocity models, offering a mechanism to mitigate non-uniqueness.

However, while these operators show great potential once trained in terms of efficiency, recent works \cite{yang_rapid_2023,zou_deep_2023} have shown that direct application of neural operator-based FWI results in noisy and unstable outputs. 
To mitigate this issue, we propose the inclusion of a novel physics-informed loss term in the FWI utilizing learned neural operators. 
Specifically, we enhance the traditional L2 norm objective function with a physics-constrained loss, evaluating the error of the learned wavefield against the wave equation, to improve the accuracy and signal-to-noise ratio of neural operator-based inversion results. 
We conduct tests on the OpenFWI datasets and realistic models to demonstrate the effectiveness of our method.

The main contributions of our work can be summarized in the following:
\begin{itemize}
    \item We utilize trained neural operators for efficient scattered wavefield solutions in an FWI framework and introduce a physics-informed loss term in FWI objectives to help improve accuracy and reduce artifacts in the inverted results.
    \item We analyze the key factors affecting the neural operator-based FWI and investigate the importance of the physics-informed loss term in the FWI to mitigate those effects.
    \item With numerical experiments on the OpenFWI and Overthrust models, we show that the proposed method can generally achieve more accurate subsurface parameter estimation compared to the vanilla approach in different scenarios.
\end{itemize} 

The remainder of this paper is organized as follows: Section \ref{theory} introduces the concepts behind our framework, including neural operator-based FWI and the physics-informed loss. Section \ref{exp} presents the numerical results in different settings, demonstrating the improvements of our framework. Finally, Sections \ref{discuss} and \ref{conclusion}  discuss and conclude our findings and future directions.

\section{Theory} 
\label{theory}
In this section, we first give a brief review of classical full waveform inversion, followed by the learned scattered wavefield solutions facilitated by the Fourier Neural Operator (FNO) for accelerating wavefield modeling. 
Then, we will introduce our physics-informed FWI using a trained neural operator for scattered wavefield solutions.

\subsection{Full waveform inversion}
\label{sec:classical_fwi}
Since utilizing the frequency-domain wavefield representation saves memory cost, which is important, compared to the time domain (a reduction in dimensionality) \cite{zou_deep_2023,huang_learned_2024}, we focus on the frequency-domain modeling in this paper.
The isotropic acoustic constant-density wave equation in the frequency domain is 
formulated as follows:
\begin{equation}
\left(\frac{\omega^2}{\mathbf{v}^2(\mathbf{x})}+\nabla^2\right) \mathbf{U}(\mathbf{x},\omega)=\mathbf{s}(\mathbf{x},\omega),
\label{equ:wave_equation}
\end{equation}
where $\omega$ is the angular frequency, $\mathbf{v}$ is the velocity, $\mathbf{U}$ is the complex frequency-domain wavefield, $\mathbf{s}$ is the source term, and $\nabla^2$ is the laplacian operator. 
Given the observation $\mathbf{d}_{obs}$, we can define the misfit function between the observed data and the simulated ones using the L2 norm:
\begin{equation}
    E(\mathbf{v}) = \sum^{N_{freq}}_{j=1}\sum^{N_{shot}}_{i=1}{\Vert \mathcal{O}(\mathcal{S}(\mathbf{v},\omega^j,s_x^i,s_z^i)) - \mathbf{d}_{obs}^{j,i}\Vert_2^2}, 
    \label{equ:classical_fwi}
\end{equation}
where $\mathcal{S}$ denotes frequency domain modeling given the source signature and velocity model, $\mathcal{O}$ denotes the source-dependent observation operator, $N_{freq}$ is the number of frequencies, and $N_{shot}$ is the number of shots. 
In the following formulation, we omit the frequency integration for notational convenience.
To minimize the misfit function and find the optimal subsurface velocity model that allows the simulated data to closely match the observed ones, we can use a gradient-based optimization method. 
The gradient of the misfit function with respect to the velocity model is given by
\begin{equation}
    \frac{\partial E}{\partial \mathbf{v}}=\operatorname{Re} \sum_{\mathrm{r}=1}^{N_{shot}}\left[\mathbf{U}^T\left(-\frac{\partial \mathbf{A}}{\partial \mathbf{v}}\right)^{\mathrm{T}} \mathbf{A}^{-1} \mathbf{R}^*\right],
    \label{equ:classic_gradient}
\end{equation}
where $Re$ denotes the real part of the complex valued term, $\mathbf{A}$ denotes the impedance matrix, and $R$ denotes the residuals at each receivers, $^*$ stands for the complex conjugation operation, $\mathbf{A}^{-1} \mathbf{R}^*$ represents the back-propagated wavefield usually obtained by the adjoint-state method.
Based on this equation, the gradient calculation process can be separated into two stages in each iteration: (1) do the forward modeling for one of the sources for the forward wavefield $\mathbf{U}$; (2) do the adjoint modeling, which is the same as the forward modeling operator for isotropic acoustic constant-density media, with the adjoint sources, which are in our case the residuals at the receiver locations. Finally, (3) multiply the two wavefields with coefficients defined by the current velocity model $\mathbf{v}$. We then obtain the gradient to update the velocity model, e.g., by $\mathbf{v}_{k+1}=\mathbf{v}_{k}-\alpha_k \frac{\partial E}{\partial \mathbf{v}}$, where $\alpha_k$ is the step length. 
It is obvious that for each iteration in FWI, we experience a large computational burden due to the repeated modeling for the forward and backward wavefields. Thus, to achieve an efficient FWI, we need to make use of a neural operator for instant wavefield simulation, and potential backpropagation, which we will introduce in the next subsection.

\subsection{Learned frequency-domain wavefield solutions}
Given a source term $\mathbf{s}$, a frequency $\omega$, and a velocity model $\mathbf{v}$, a neural network $f$ with parameters $\theta$ can learn to map from $\{\mathbf{v},\mathbf{s},\omega\}$ to the wavefield $\mathbf{U}$.
We will consider here the 2D case. However, the implementation is equally valid for 3D.
Then, we can replace the modeling operator $\mathcal{S}(\mathbf{v},\omega,s_x,s_z)$ with $f_{\theta}(\mathbf{v},\omega,s_x,s_z)$.
However, when predicting wavefield solutions, the source location $(s_x, s_z)$ is treated as a binary mask, and the $\omega$ is often used by the partial channel with a constant value, yielding suboptimal representation for a neural network to extract the useful features.
As suggested by \cite{huang_learned_2024}, embedding the information of the source location and frequency in the background wavefield $\mathbf{U}_0$ could provide a compact and flexible representation of the input. 
Trained over multiple frequencies and source locations, we learn to map with $\{\mathbf{v},\mathbf{U}_0\}$ as input, in which the background wavefield is given analytically for a homogeneous background velocity (fixed for all training inference samples), $\mathbf{v}_0$, by
\begin{equation}
\mathbf{U}_0(x, z)=\frac{i}{4} \boldsymbol{H}_0^{(2)}\left(\omega \sqrt{\frac{\left\{\left(x-s_x\right)^2+\left(z-s_z\right)^2\right\}}{\mathbf{v}_0^2}}\right),
\label{equ:background}
\end{equation}
where $\boldsymbol{H}_0^{(2)}$ is the zero-order Hankel function of the second kind, $(s_x,s_z)$ define the source location.
To further enhance the performance of the neural operator for surrogate modeling of the wave equation, instead of predicting the whole wavefield, we 
choose to predict the residuals of the PDE solutions \cite{wandel_teaching_2021}, that is the scattered wavefield $\delta\mathbf{U}=\mathbf{U}-\mathbf{U}_0$, satisfying
\begin{equation}
\frac{\omega^2}{\mathbf{v}^2} \delta \mathbf{U}+\nabla^2 \delta \mathbf{U}+\omega^2\left(\frac{1}{\mathbf{v}^2}-\frac{1}{\mathbf{v}_0^2}\right) \mathbf{U}_0=0.
\label{equ:scattered_equation}
\end{equation}

Our primary emphasis in this paper is on the inversion process, thus, we will use a supervised mechanism for the training of the neural operator. 
Adhering to the conventional supervised learning paradigm, we randomly sample source locations and frequencies from a uniform distribution over a specific range, analytically calculate the background wavefield with a fixed constant background velocity model, and solve for the corresponding scattered wavefield using a finite difference numerical method on randomly sampled velocity models from within the training dataset. 
We use absorbing boundary conditions at all boundaries to avoid artificial reflections. While this setup excludes free surface-related wave phenomena, our method does not rely on any particular boundary condition and can be extended to free surface cases with appropriate training data.
We use these samples to train a neural network $f_\theta$ using a mean squared error loss
\begin{equation}
    L(\theta) = \frac{1}{N_{s}}\sum^{N_{s}}_{i=1}{\Vert f_\theta(\mathbf{v}^i,\mathbf{U}_0^i)-\delta\mathbf{U}^i\Vert_2^2},
    \label{equ:fno_loss}
\end{equation}
where $N_s$ is the number of training samples. For the choice of neural network in this paper, we use FNO \cite{li_fourier_2020}, which has been introduced as a strong benchmark for the parametric PDE solutions, to learn the wavefield solutions. 

\subsection{Physics-constrained loss for neural operator-based full waveform inversion}
Inserting the neural operator-based wavefield solution into equation~\ref{equ:classical_fwi}, the neural operator-based FWI using an L2 norm objective function is formulated as follows:
\begin{equation}
    E(\mathbf{v}) = {\Vert \mathcal{O}(f_{\theta}(\mathbf{v},\mathbf{U}_0)+\mathbf{U}_0) - \mathbf{d}_{obs}\Vert_2^2}.
    \label{sec8:l2norm}
\end{equation}
Besides the efficiency of the neural operator-based wavefield modeling, another benefit of neural operator-based FWI is the flexibility of the gradient calculation.
Different from the numerical way in conventional FWI relying on the adjoint method, which requires the formulation of the adjoint modeling operator and derivation of the gradient in terms of different FWI parameterization, the neural operator-based FWI can alternatively produce the gradient of the misfit function with respect to the input velocity model by automatic differentiation \cite{baydin_automatic_2018}.
Even though combined with other regularization terms, e.g., total variation regularization term, which is often solved by a fast iterative shrinkage-thresholding algorithm, 
the neural operator-based FWI can easily handle the gradient calculation.

Although efficient and flexible, the neural operator-based FWI often results in noisy subsurface parameter inversion results, which are not physically explainable.
To obtain more continuous and accurate inverted results, we need to eliminate such non-physical artifacts in the inversion process.
In other words, we hope the data (seismic wavefield) and velocity models obey the physical laws, which is similar to the objective of the wavefield reconstruction inversion \cite{van_leeuwen_mitigating_2013,leeuwen_penalty_2015}.
Hence, we propose to utilize a physics-constrained loss to regularize the inversion process. 
Specifically, we add it as a soft constraint given by the governing equation \ref{equ:scattered_equation}, resulting in a new objective function:
% \begin{equation}
\begin{dmath}
    E(\mathbf{v}) = \frac{1}{N_o}\Vert \mathcal{O}(f_{\theta}(\mathbf{v},\mathbf{U}_0)+\mathbf{U}_0) -\mathbf{d}_{obs}\Vert_2^2 + 
    \allowbreak
    \frac{\lambda}{N} \sum_{i=1}^N\left\Vert \frac{\omega^2}{(\mathbf{v}^i)^2} f_\theta(\mathbf{v}^i,\mathbf{U}_0^i)+
    \allowbreak
    \nabla^2 f_\theta(\mathbf{v}^i,\mathbf{U}_0^i)+ 
(\frac{\omega^2}{(\mathbf{v}^i)^2}-\frac{\omega^2}{(\mathbf{v}_0^i)^2})\mathbf{U}_0^i\right\Vert_2^2,
    \label{sec8:l2normnew}
% \end{equation}
\end{dmath}
where $\lambda$ is a hyperparameter balancing the contribution of the loss terms and chosen by making both terms of comparable scale, $N_o$ is the number of observation points, and $N$ is the number of points in the spatial domain.
The second term is typically used in physics-informed machine learning as a loss function to optimize the neural network to converge to a PDE-based solution (physically accurate), in which the velocity is fixed while the wavefield is the fitting target to optimize the neural network. 
However, in the proposed method, our objective is velocity. 
At the same time, the wavefield is predicted from a frozen (pre-trained model with fixed weights and biases during the FWI) neural operator. 
This is different from the PINN-based inversion, where the PDE loss function is used to optimize the velocity as well as the wavefield modeling \cite{Song2022waveinv,schuster_review_2024}.

\begin{figure}[!htb]
    \centering
    \includegraphics[width=1.0\columnwidth]{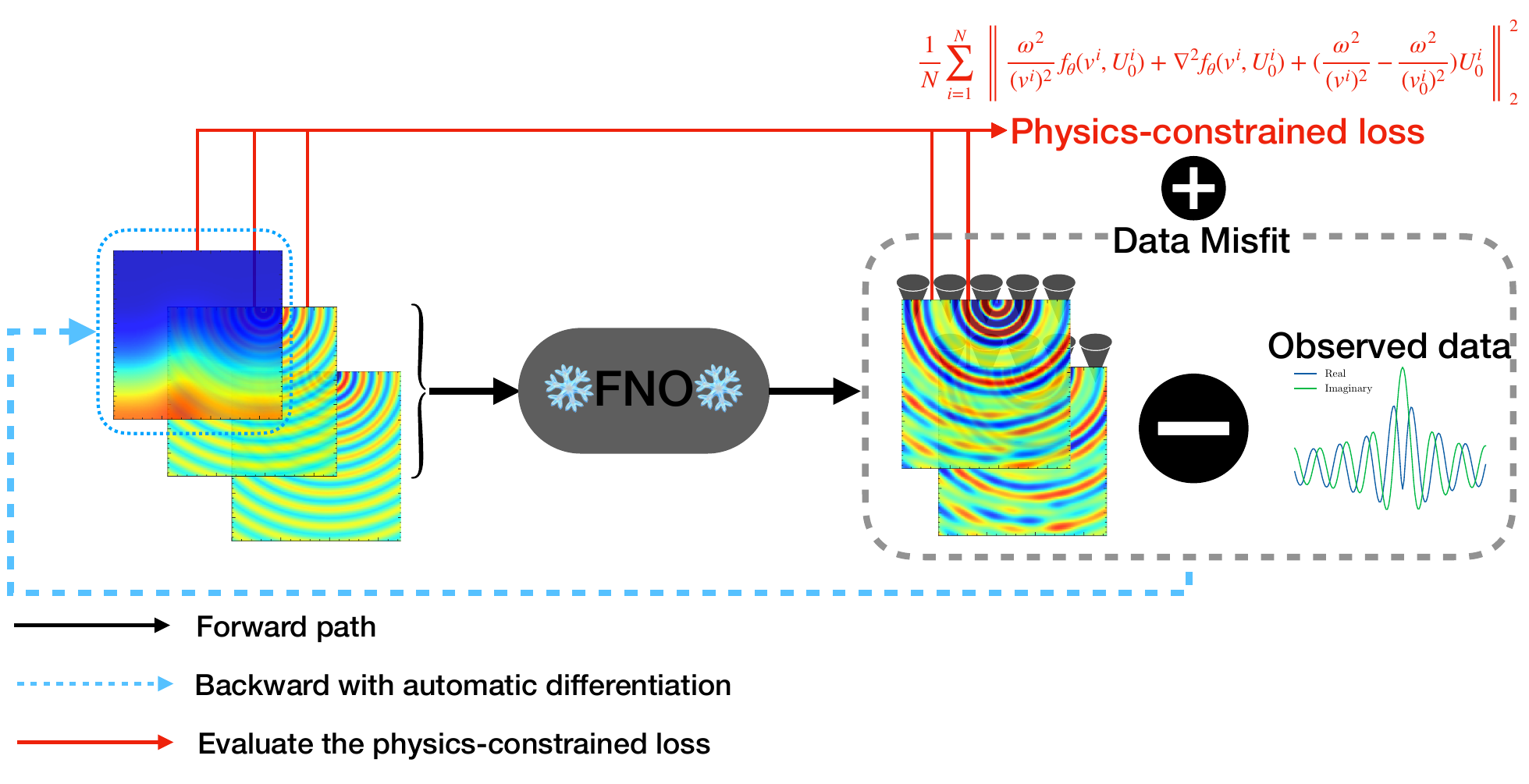}
    \caption{A diagram of the physics-informed FWI using trained neural operators based wavefield solutions. 'FNO' denotes the learned simulation engine for the wavefield solutions, whose parameters are frozen during FWI. 
    The red line denotes the path to evaluate the physics-constrained loss, and the blue dashed line denotes the backpropagation process with automatic differentiation to update the input velocity.
    The rest, which excludes the path to evaluate the physics-constrained loss, is the vanilla neural operator-based FWI using the L2 norm objective function.}
    \label{fig:diagram}
\end{figure}

The whole framework of the proposed physics-informed FWI using learned wavefield solutions is shown in Figure~\ref{fig:diagram}. 
Before the inversion process, we need to train a neural operator with a large number of velocity model samples. Hopefully, these samples include our prior knowledge of the subsurface under investigation, or at least our knowledge that velocity often increases with depth and the Earth is generally layered. The labels for the training are the corresponding numerical wavefield solutions for these models based on equation~\ref{equ:fno_loss}. 
After the training, we freeze the weights of the neural operator during the inversion process. 
Recall that what the neural operator learns are the solutions of parametric PDEs, specifically for an in-distribution velocity model.
Then, given the observed data, we calculate the misfit between them and predicted ones from the neural operator for an initial velocity model. 
For the conventional neural operator-based FWI, we directly backward propagate the misfit using automatic differentiation and update the initial velocity model.
For the proposed method, before the backpropagation process, we evaluate the PDE residuals given by the second term in equation~\ref{sec8:l2normnew} with the predicted wavefield and the velocity in the current loop used for the wavefield prediction.
After that, we backpropagate the new loss values to update the input velocity in the current loop directly. 
By iteratively doing so, we can minimize the loss function and hopefully converge to a credible velocity model.

\subsection{Details of the backbone network and training configuration}
\begin{figure}[!htb]
    \centering
    \includegraphics[width=1.0\columnwidth]{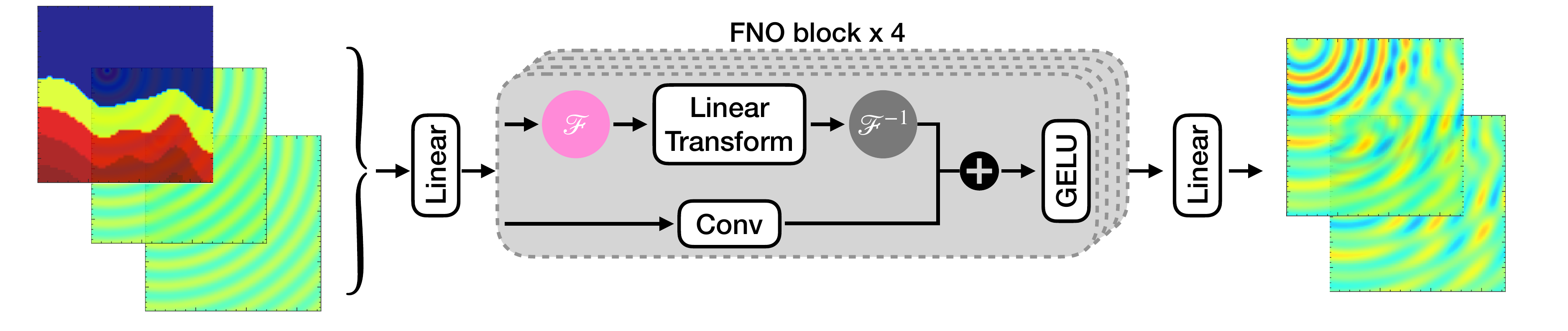}
    \caption{The architecture of the FNO for learned wavefield solutions.}
    \label{fig:diagram_fno}
\end{figure}

In this subsection, we will share the details of the network we use and the training configuration. 
As shown in Figure~\ref{fig:diagram_fno}, there are four FNO blocks in general.
The input to the network comprises three channels, which include the velocity model and real and imaginary parts of the background wavefield. The output of the network comprises two channels, which include the real and imaginary parts of the scattered wavefield. 
Before inputting those three channels into the FNO blocks, we use the linear point-wise feedforward layer, transforming the original three channels into a high-dimensional space of 128 channels. 
Then, this high-dimensional tensor will be fed into a stack of four sequential FNO blocks.
In each FNO block, the data are transformed into the wavenumber-domain with Fast Fourier transform (FFT, denoted as $\mathcal{F}$).
For efficiency, we usually do not process the full coefficients (wavenumbers) of the transformed tensors and typically cut the frequency mode at a certain number, which is 48 in both $x$ and $z$ directions.
Then, we apply the linear transformation, which is a complex multiplication in the channel space, followed by the inverse FFT ($\mathcal{F}$) to transform them back to the spatial domain.
There is also a residual connection, in which we do a simple convolutional operation with a kernel size of $1\times1$ to perform the channel mixing and then add them with the results after inverse FFT.
Finally, to provide the neural network non-linear representation capabilities, a non-linear activation function, specifically the GELU \cite{hendrycks_gaussian_2016} activation function here, is employed.
After four sequential blocks, there is one additional linear point-wise projection layer transforming the high-dimensional data to the target spatial space.

All following experiments are performed on an NVIDIA A100 GPU. During the training of the FNO for predicting the wavefield solution, we used an Adam optimizer with a fixed learning rate. 
In the FWI stage, we also use an Adam optimizer to update the velocity model with a fixed learning rate. 

\section{Numerical experiments}
\label{exp}
In this section, we will conduct tests on the “CurveVelA” class velocity model set from the OpenFWI dataset \cite{deng_openfwi_2022}. This dataset features velocities with curvatures that increase with depth. 
We also showcase the performance of the proposed method on the Overthrust model.
Our aim is to investigate the source of the noisy reconstruction using neural operator-based FWI and demonstrate the importance and effectiveness of adding a physics-constrained loss term for improving the inverted results. 
Subsequently, considering the different scenarios that the real application of neural operator-based FWI might face, we investigate the issues that affect the inversion results, and the improvement performance of the proposed method. 

\subsection{"Ideal tests" on "CurveVelA" velocity models}
\label{ideal_curvea}
\begin{figure}[!htb]
    \centering
    \includegraphics[width=0.8\columnwidth]{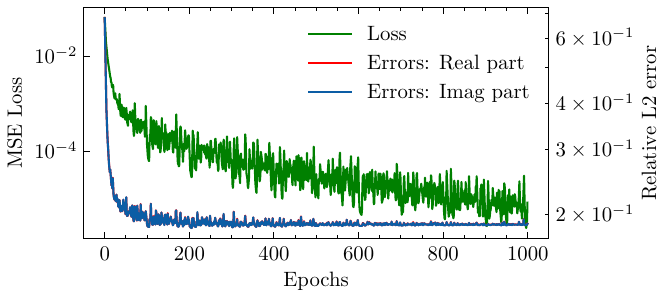}
    \caption{The training loss history for the learned wavefield solutions and the relative errors for the real and imaginary parts of the wavefield on the validation set.}
    \label{fig:training_loss_curvea}
\end{figure}

In this subsection, we first assess the performance of the neural operator-based FWI, which employs automatic differentiation to iteratively refine the velocity model.  
The velocity model is first upsampled to a resolution of 139$\times$139, reducing the spatial interval to 0.0125 $km$.
To train the FNO for the wavefield simulation in our FWI, we created a dataset comprising 9,000 samples, each characterized by a specific source location, frequency, and velocity model.
For each training sample, the source locations were randomly sampled from just below (depth 0.025 km) the surface boundary to simulate conventional acquisition geometries, and the frequency was chosen randomly from between 3 to 15 Hz. 
We configured the training process with a learning rate of 0.001 and a batch size of 128 and trained for a total of 1,000 epochs. 
The training loss history is shown in Figure~\ref{fig:training_loss_curvea}. 
The performance on the validation set demonstrates that the neural operator converges well.   

\begin{figure}[!htb]
    \centering
    \includegraphics[width=1.0\columnwidth]{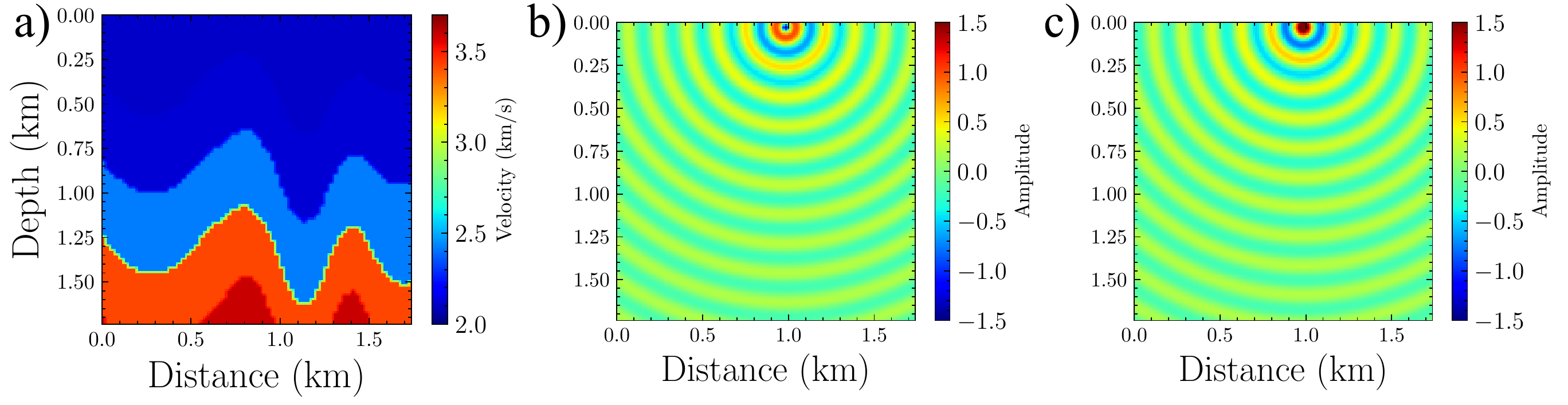}
    \caption{From left to right are the true velocity (a), corresponding real (b) and imaginary (c) parts of the background wavefield due to a single source at $(0.98 km, 0.025 km)$ with a frequency of 8.8 Hz .}
    \label{fig:fwi_curvea_true_u0}
\end{figure}

After training, we froze the FNO parameters and used the neural operator in FWI.
In the first subsection, we aim to analyze the mechanism behind the noisy inversion and the compensation of physics-informed loss.
Hence, we consider the "ideal tests", where the accuracy of the forward modeling by the neural network on the target velocity is good enough. 
We randomly picked one sample $(\mathbf{v}, \omega, s_x, s_z)$, shown in Figure~\ref{fig:fwi_curvea_true_u0}, which is not seen during the training, but its velocity is seen during the training and there are similar shots on the same velocity model in the training datasets.
This is to guarantee that the neural operator-based forward modeling is accurate, at least on the target velocity model and source signature, so-called "ideal tests".
To simplify the scenario and test the performance of the neural operator-based FWI, we choose only one source and a single frequency for the FWI. 

\begin{figure}[!htb]
    \centering
    \includegraphics[width=1.0\columnwidth]{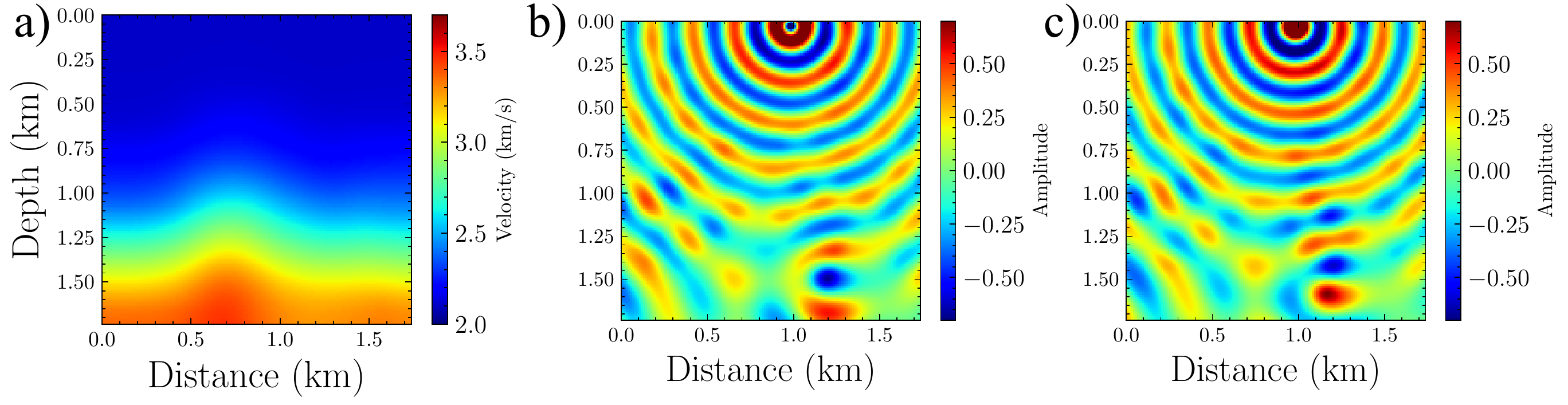}
    \caption{From left to right are the smoothed initial velocity, corresponding real and imaginary parts of the observed wavefield due to a single source at $(0.98 km, 0.025 km)$ with a frequency of 8.8 Hz .}
    \label{fig:fwi_curvea_initial_obs}
\end{figure}

\begin{figure}[!htb]
    \centering
    \includegraphics[width=1.0\columnwidth]{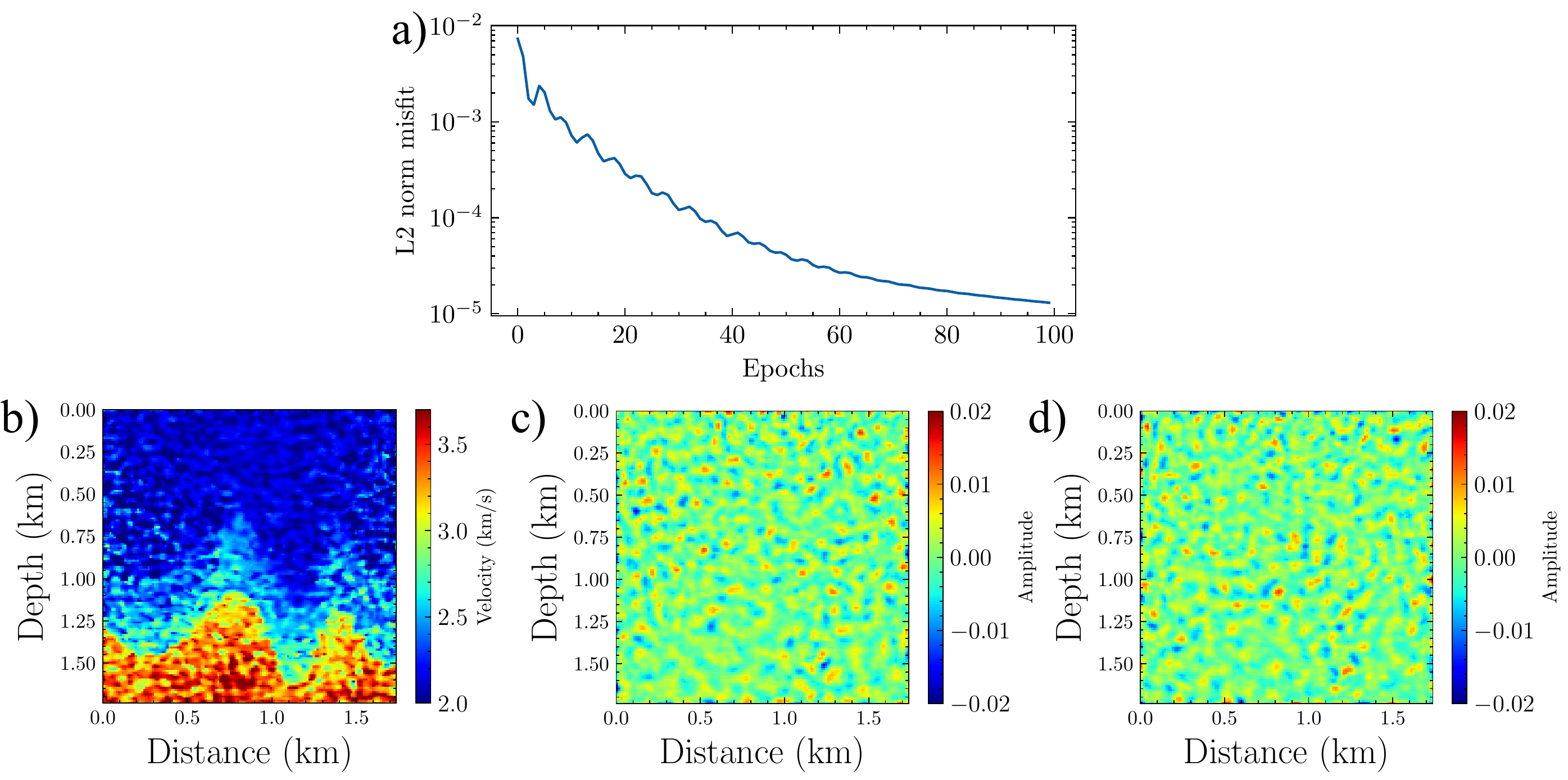}
    \caption{The inverted result using full space observations. a) the L2 objective function along the iterations; b) the inverted result after 100 iterations; The real c) and imaginary d) parts of the residuals between the observed wavefields and predicted ones using the inverted velocity model.}
    \label{fig:fwi_curvea_misfit_inv}
\end{figure}

To test the approach, we first applied FWI, assuming observations over the entire spatial domain, as shown in Figures~\ref{fig:fwi_curvea_initial_obs}b and \ref{fig:fwi_curvea_initial_obs}c. We smoothed the ground truth velocity model (Figure~\ref{fig:fwi_curvea_true_u0}a) with a Gaussian filter to serve as the initial model (Figure~\ref{fig:fwi_curvea_initial_obs}a). The inverted result after 100 iterations, using only the L2 norm objective function (Equation~\ref{equ:classical_fwi}) with an Adam optimizer and a learning rate of 0.05, is shown in Figure~\ref{fig:fwi_curvea_misfit_inv}.
We observe that the FWI misfit function (Figure~\ref{fig:fwi_curvea_misfit_inv}a) decreases significantly, and the prediction based on the final inverted results is close to the observed full wavefield with minimal errors. However, the final inverted result (Figure~\ref{fig:fwi_curvea_misfit_inv}b) is quite noisy, even though the general shape and boundary between different layers can be clearly observed.
We simulated the wavefield using this inverted result and obtained the residuals between the predicted real and imaginary parts and the observed ones. As shown in Figures~\ref{fig:fwi_curvea_misfit_inv}c and d, the residuals are very low in amplitude. This is not an ideal situation for neural operator-based inversion, as it indicates that the inverted result will not be reliable, even though the observed data is well-fitted.

\begin{figure}[!htb]
    \centering
    \includegraphics[width=1.0\columnwidth]{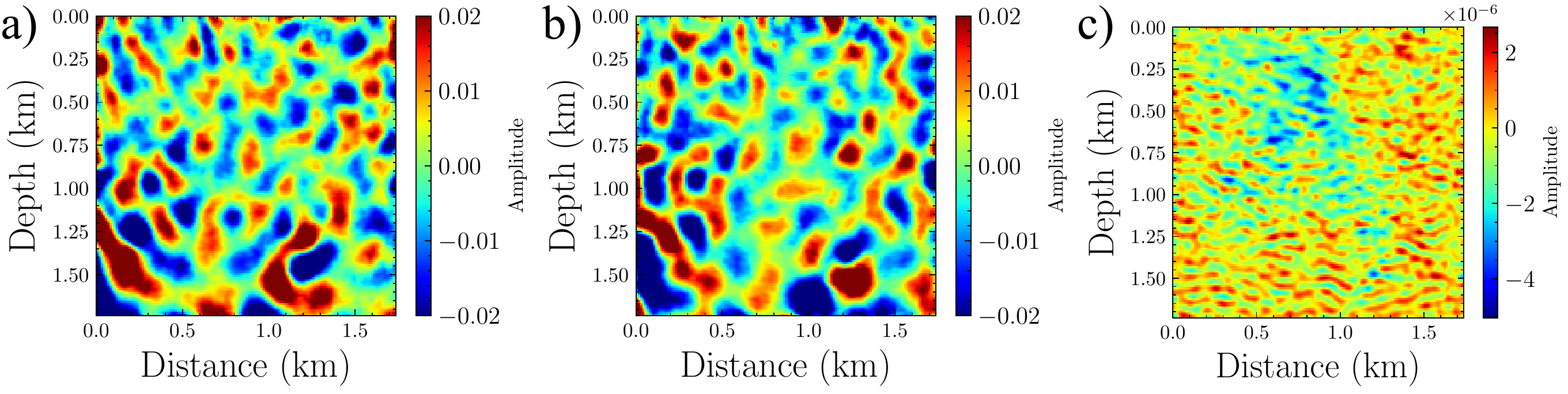}
    \caption{The real a) and imaginary b) parts of the residuals between the observed wavefields and predicted ones using the ground truth velocity model. c) is the gradient estimated by the automatic differentiation.}
    \label{fig:fwi_curvea_grad_from_true}
\end{figure}

\begin{figure}[!htb]
    \centering
    \includegraphics[width=1.0\columnwidth]{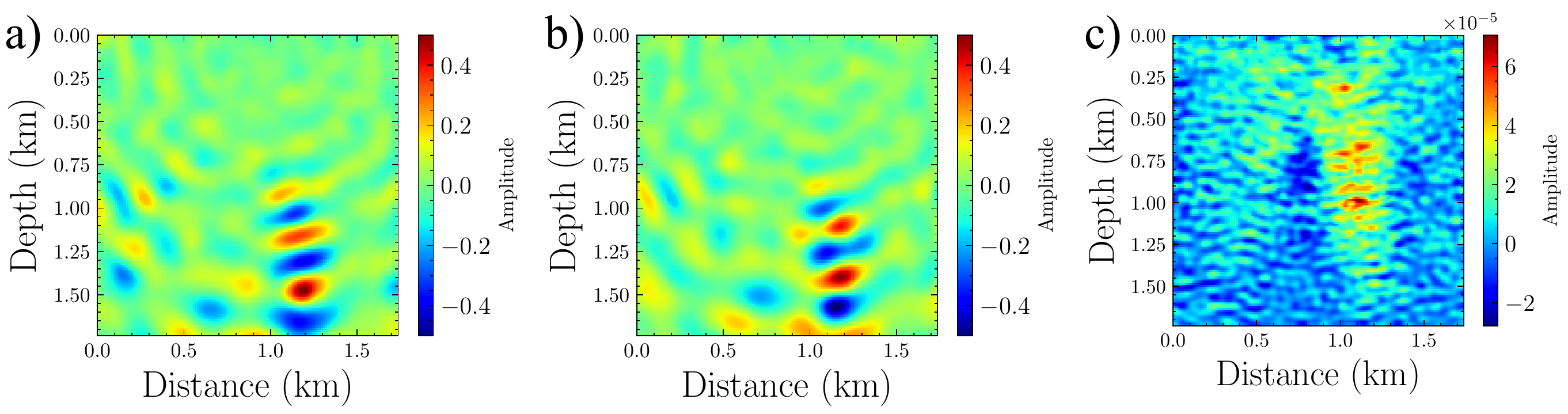}
    \caption{The real a) and imaginary b) parts of the residuals between the observed wavefields and predicted ones using the initial model (Figure~\ref{fig:fwi_curvea_initial_obs}a). c) is the gradient estimated by the automatic differentiation.}
    \label{fig:fwi_curvea_grad_from_init}
\end{figure}

\begin{figure}[!htb]
    \centering
    \includegraphics[width=0.7\columnwidth]{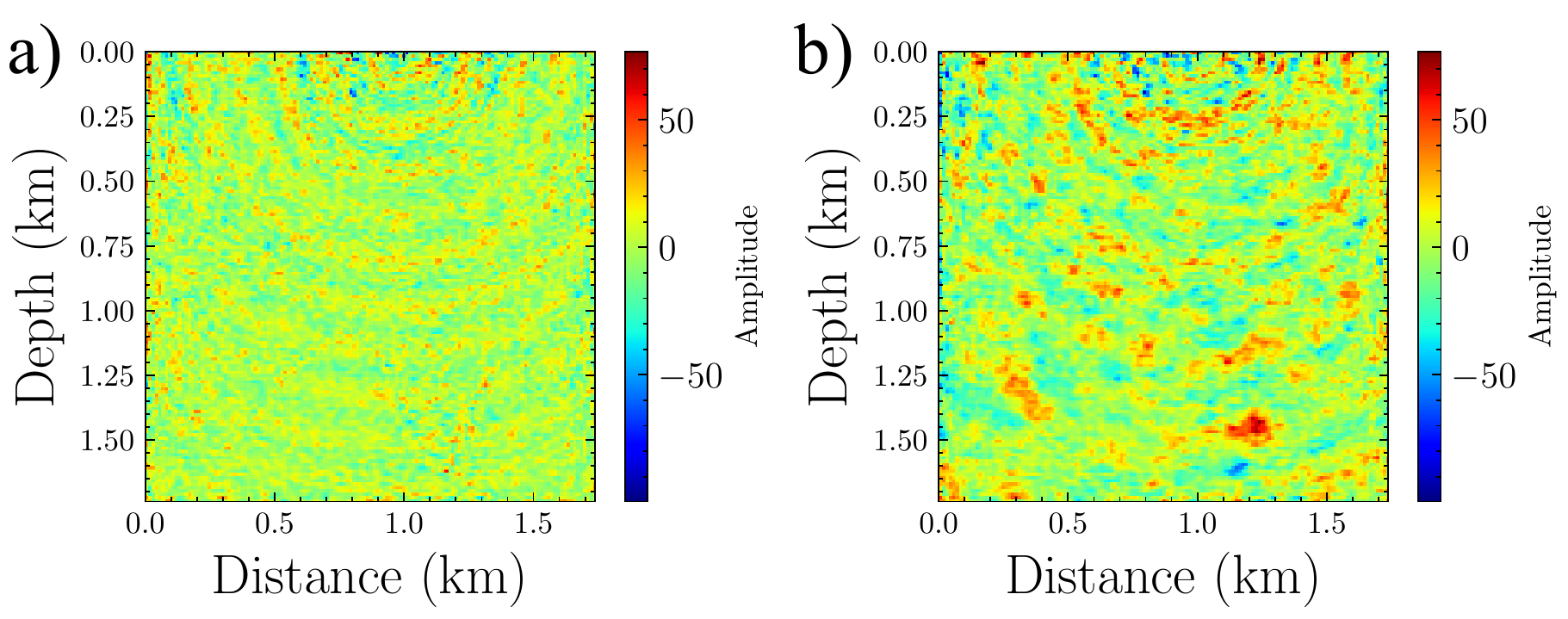}
    \caption{The PDE (physics-constrained loss) residuals on a) ground truth and that b) on the initial model (Figure~\ref{fig:fwi_curvea_initial_obs}a) predictions.}
    \label{fig:fwi_curvea_pde_from_true_init}
\end{figure}

We dive more into the mechanism of this phenomenon by estimating the gradient of the FWI when the inverted result is the same as the ground truth. 
This is shown in Figure~\ref{fig:fwi_curvea_grad_from_true}.
Although with the ground truth velocity model used as input, we end up with small residuals between the predicted wavefield using the neural operator and the observed wavefields. 
We backpropagate those residuals to calculate the corresponding gradient, shown in Figure~\ref{fig:fwi_curvea_grad_from_true}c.
It indicates that when the inverted velocity model reaches the ground truth, the inherent predicted errors from the surrogate modeling will be propagated back, resulting in noisy gradients, which is one of the reasons for the noisy inverted result.
This should happen at the later stage of the inversion process.
However, we found that at the early stage, for example, the first iteration of the FWI process, where the residuals between the simulated wavefield and observed ones have clear information related to the subsurface velocity model difference, the gradient is still noisy (shown in Figrue~\ref{fig:fwi_curvea_grad_from_init}c), which is different from conventional FWI, in which the gradient of the first iteration should mainly reflect the structure of the subsurface model. 
The neural operator will accumulate the noise, yielding the final noisy inverted result.
This shows that beyond the inherent accuracy issue of the surrogate modeling, the generalization to any perturbation in the velocity model is another issue causing the noisy gradient calculation. Specifically, the neural operator tends to encode features associated with sharp velocity contrasts, as it was primarily trained on high-resolution models. This bias limits its ability to accurately represent wavefields for smoother or perturbed velocity models encountered during early inversion stages.
We plot the PDE residuals for the real parts of the scattered wavefield for the prediction using equation~\ref{equ:scattered_equation}, in Figure~\ref{fig:fwi_curvea_pde_from_true_init}. 
Clearly, the PDE loss is higher for the initial smooth velocity model (Figure~\ref{fig:fwi_curvea_pde_from_true_init} b) compared to that for the true velocity (Figure~\ref{fig:fwi_curvea_pde_from_true_init} a). This is a direct consequence of this ground truth model coming from the trained distribution, while the smooth model did not. 
So, the ground truth velocity model obeys the physics that was employed in the numerical simulation of the wavefields for these sharp velocity models. On the other hand, wavefield simulations for smooth velocity models were not embedded in the neural operator training.
This motivates us to utilize the physics-constrained loss during the inversion process to compensate for the update.

\begin{figure}[!htb]
    \centering
    \includegraphics[width=1.0\columnwidth]{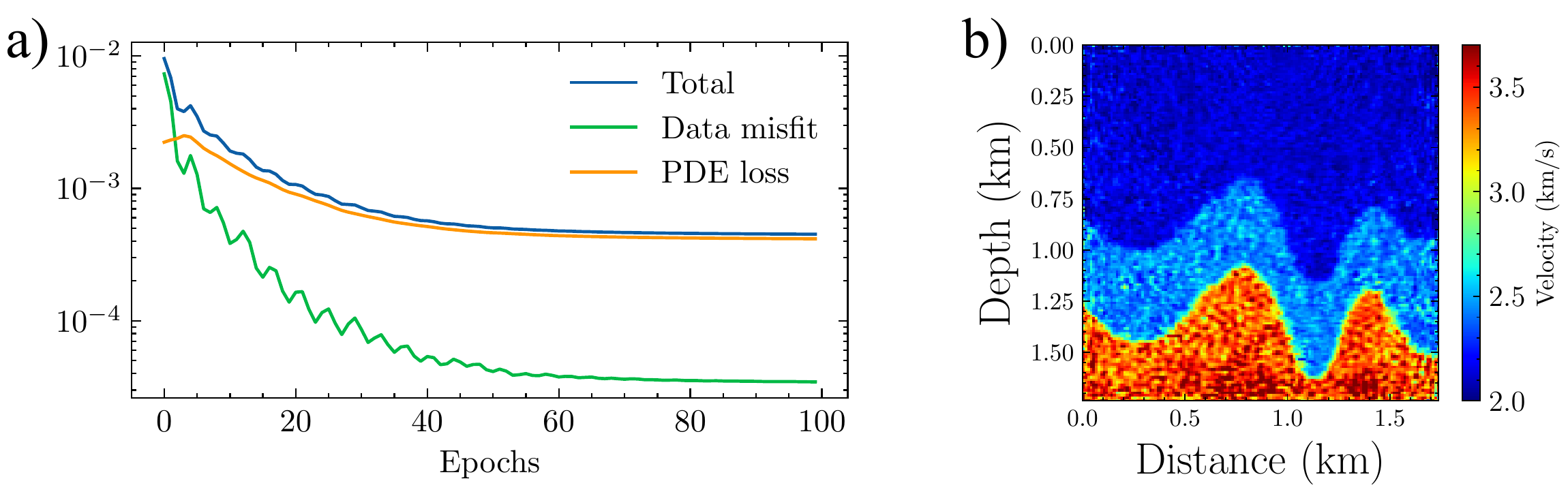}
    \caption{The convergence history (a) of physics-constrained FWI and the inverted result starting from the initial model (Figure~\ref{fig:fwi_curvea_initial_obs}a).}
    \label{fig:fwi_curvea_l2_pde_inv}
\end{figure}

As a result, in addition to the conventional neural operator-based FWI pipeline, our method evaluates the PDE residuals as an additional misfit term to update the velocity model in each iteration. 
Here we use the same optimizer and the same learning rate as those used in the conventional neural operator-based FWI, and set $\lambda$ to 1.9e-6.
The convergence history and inverted result are shown in Figure~\ref{fig:fwi_curvea_l2_pde_inv}. With the constraint of the physics-informed loss term, the final inverted result is much more accurate compared to the one using conventional neural operator-based FWI, which uses only the L2 objective function. The proposed method not only reduced the noise and artifacts but also recovered the boundaries of the layers. 
Proper selection of the hyperparameter $\lambda$ plays a key role in balancing the data loss and the physics-constrained PDE loss during inversion. To evaluate its sensitivity, we tested five different values: 1e-7, 1e-6, 1e-5, 1e-4, and 1e-3, representing different scales. The corresponding inversion accuracies (in terms of relative errors of the velocity models compared to the ground truth, which is equal to the norm of the errors divided by the norm of the ground truth velocity model) are 0.0489, 0.0354, 0.0420, 0.0476, and 0.0488, respectively. These results suggest that the performance is relatively robust within a reasonable range of $\lambda$, and that choosing $\lambda$ such that the data loss and PDE loss are of comparable magnitude is crucial for achieving improved inversion quality.

\begin{figure*}[!htb]
    \centering
    \includegraphics[width=2.0\columnwidth]{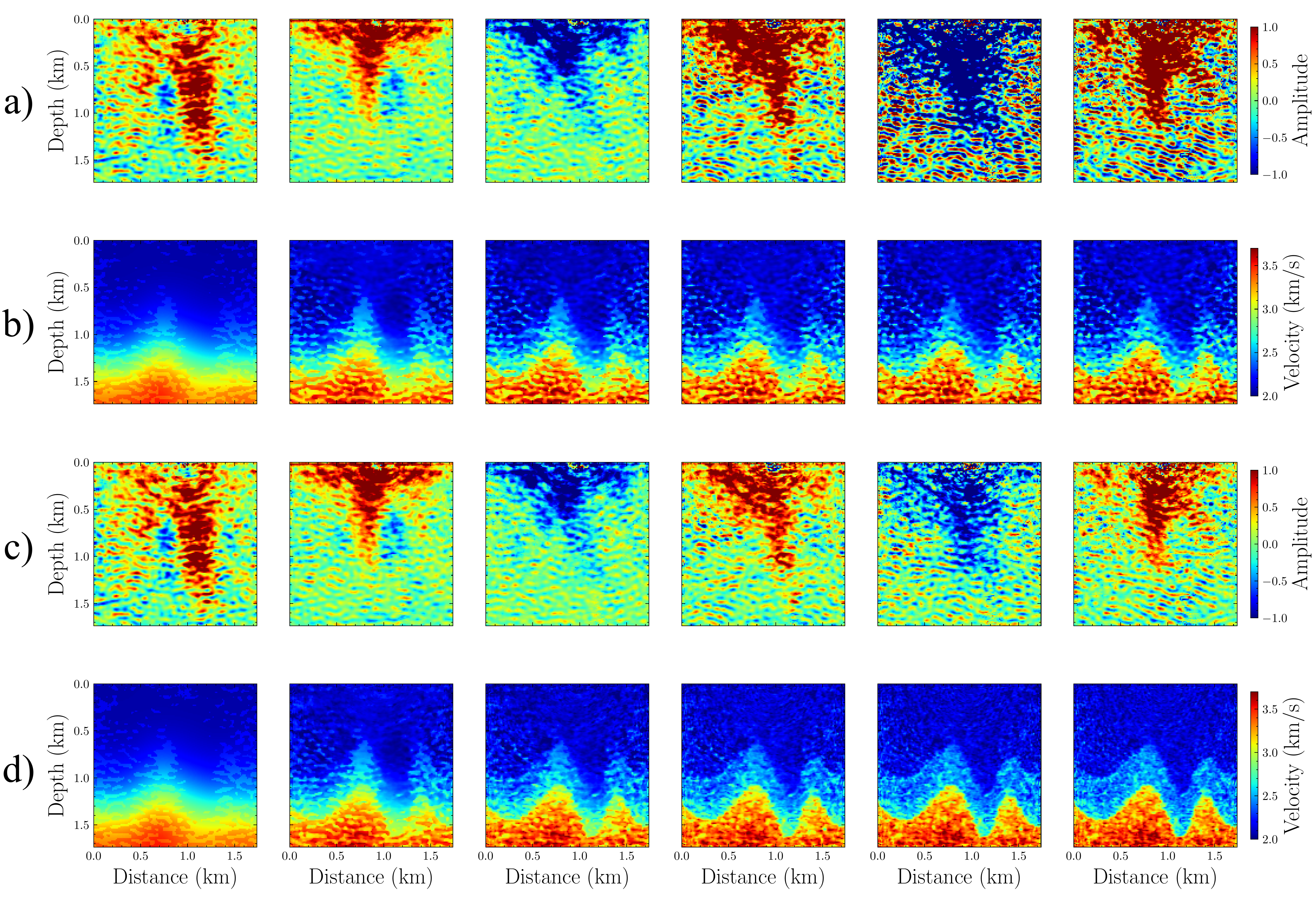}
    \caption{The comparison of the gradients (a, c) and updated velocity models (b, d) using the conventional neural operator based FWI (a, b) and that using the proposed method (c, d). Each columns represent the results at different iterations, which are 0, 4, 8, 12, 16, 20 from left to right.}
    \label{fig:fwi_curvea_l2_pde_inv_vs_l2}
\end{figure*}

To further understand the importance and effectiveness of the physics-constrained loss for FWI, we show the gradients within the first 21 iterations and the corresponding updated velocities. We specifically plot the gradient and update for every 4 iterations in Figure~\ref{fig:fwi_curvea_l2_pde_inv_vs_l2}. It is evident that at the early stage, because the loss function is still dominated by the data misfit and the PDE loss has not decreased, the gradient and the updated velocities of the proposed method are almost the same as those using the conventional method. However, after 9 iterations, we can easily observe the compensation from the PDE loss term, which eliminates some noise and artifacts in the gradient (Figure~\ref{fig:fwi_curvea_l2_pde_inv_vs_l2}c, column 4), resulting in a better update (Figure~\ref{fig:fwi_curvea_l2_pde_inv_vs_l2}d, column 4). These compensation effects become more apparent in subsequent iterations. The timing of this phenomenon coincides with the decrease of the PDE loss shown in Figure~\ref{fig:fwi_curvea_l2_pde_inv}, demonstrating the effectiveness and importance of the proposed method.

\begin{figure}[!htb]
    \centering
    \includegraphics[width=1.0\columnwidth]{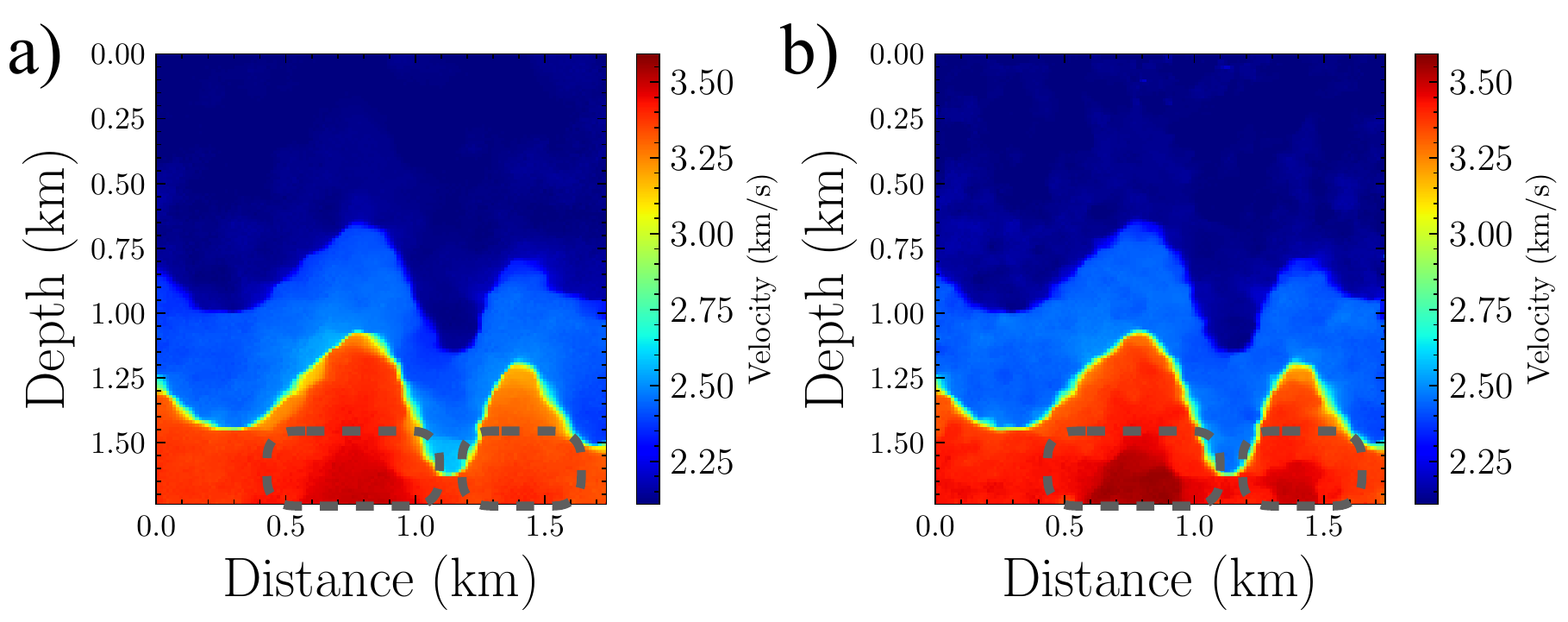}
    \caption{
    The comparison of inverted results with TV regularization using the conventional neural operator-based FWI (a) and that using our proposed method (b). The dashed gray box highlights the main differences.}
    \label{fig:fwi_curvea_l2_pde_t v_inv_vs_l2_tv}
\end{figure}

To further improve the signal-to-noise ratio of the inverted results, we applied the commonly used Total Variation (TV) regularization.
The results using the conventional neural operator-based FWI and our proposed method are shown in Figure~\ref{fig:fwi_curvea_l2_pde_t v_inv_vs_l2_tv}. 
Although the TV regularization improved the inverted result for both methods, we can clearly observe the benefit of the PDE loss in such a setting, denoted by the Gray dashed box. 
Our method successfully reconstructs the deeper part of the velocity model and the corresponding boundary, while the conventional neural operator-based FWI fails. However, as a reminder, we assume recording over the whole domain for these tests.
To quantify the inverted results, we calculate the relative errors of the velocity model compared to the ground truth. 
The error of our proposed method is 0.0152, lower than that of the conventional neural operator method, which is 0.0182.

\subsection{"Ideal tests" on Overthrust model}
\label{sec:ideal-ove}
To further test the effectiveness of the proposed method, we also conducted tests on selected velocities from the Overthrust model, which are more realistic. 
Similar to the experiment setting above, we perform the inversion using one source and a single frequency and compare the results of the standard (vanilla) implementation with our approach, which incorporates a physics-constrained loss term. 
To thoroughly evaluate the performance of our method, we employ two distinct methods of wavefield recording: one, like the previous example, recording the wavefield across the entire spatial domain and another focusing solely on the area near the surface. 
These tests allow us to examine the impact of our physics-constrained loss term on the inversion accuracy for both a perfect observation and in scenarios where data might be limited to surface recordings.

Again, before employing FWI, we trained our FNO architecture for the wavefield simulation on the distribution of the Overthrust model. 
We created a dataset comprising 5,400 samples, each characterized by a random source location on the surface, frequency, and velocity. 
The velocity model is discretized at a resolution of 256$\times$256 and a spatial interval of 0.025 km. 
The training configuration and sampling of the source locations are the same as above, but the frequency was chosen from between 3 to 21 Hz, encompassing the primary frequency band often utilized in practical FWI.
Similar to the "ideal tests" on "CurveVelA" velocity models, to ensure the accuracy of the predicted wavefield on the target velocity models and focus more on the realistic model and imperfect observations, we invert an in-distribution sample, of which the velocity is seen during the training and the source configuration is unseen but similar to the training shots. 
The source is located at the location of 3.69 km and with a frequency of 10.0 Hz. 

\begin{figure}
    \centering
    \includegraphics[width=1.0\columnwidth]{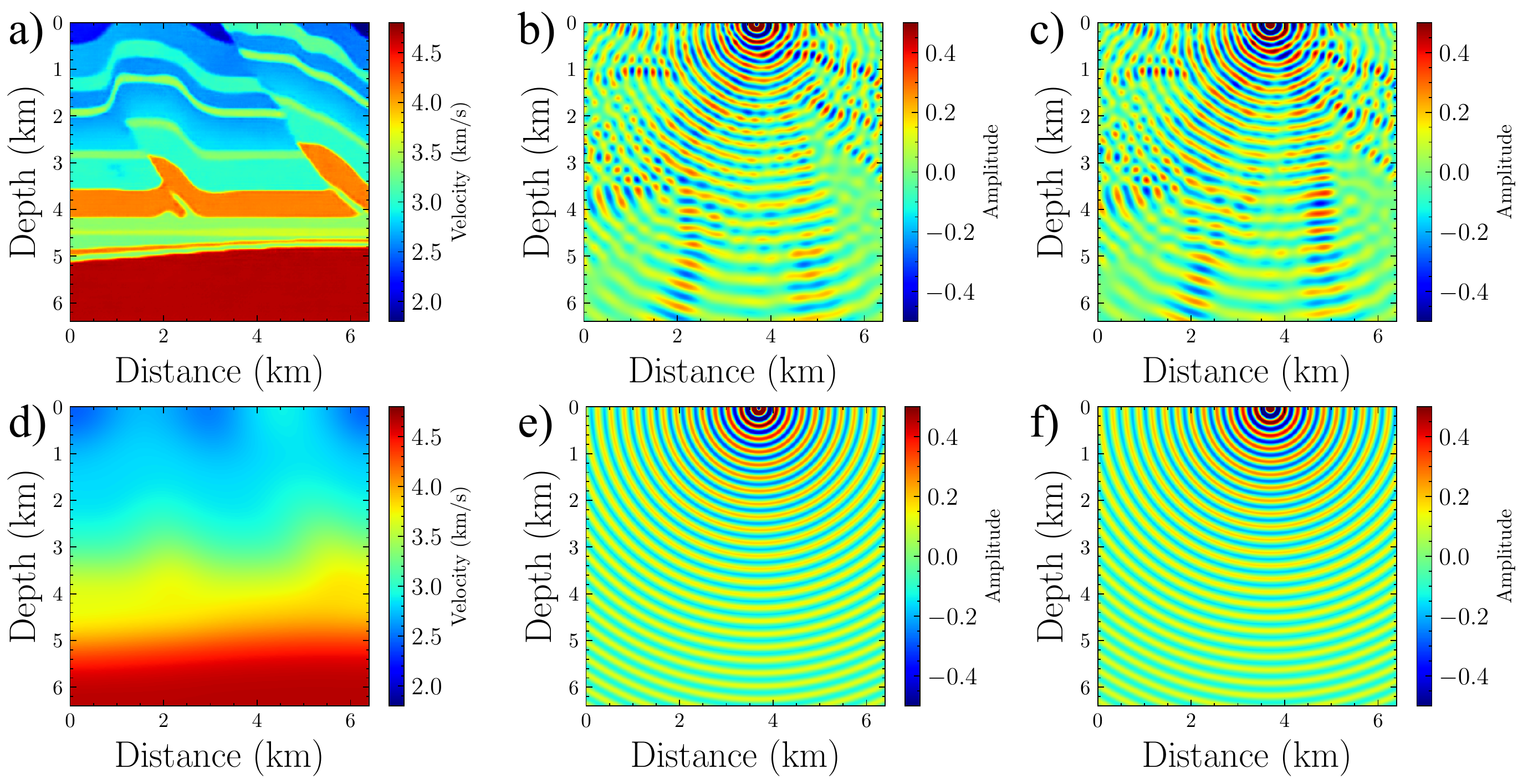}
    \caption{The true velocity (a) and the real (b) and imaginary (c) parts of the observed wavefield with a frequency of 10 Hz using numerical finite-difference simulation. 
    (d) is the initial model used in the FWI, and (e) and (f) show the real and imaginary parts of the analytical background wavefield, which is fixed throughout the inversion.}
    \label{fig:fwi_seg_init_true}
\end{figure}

\begin{figure}
    \centering
    \includegraphics[width=1.0\columnwidth]{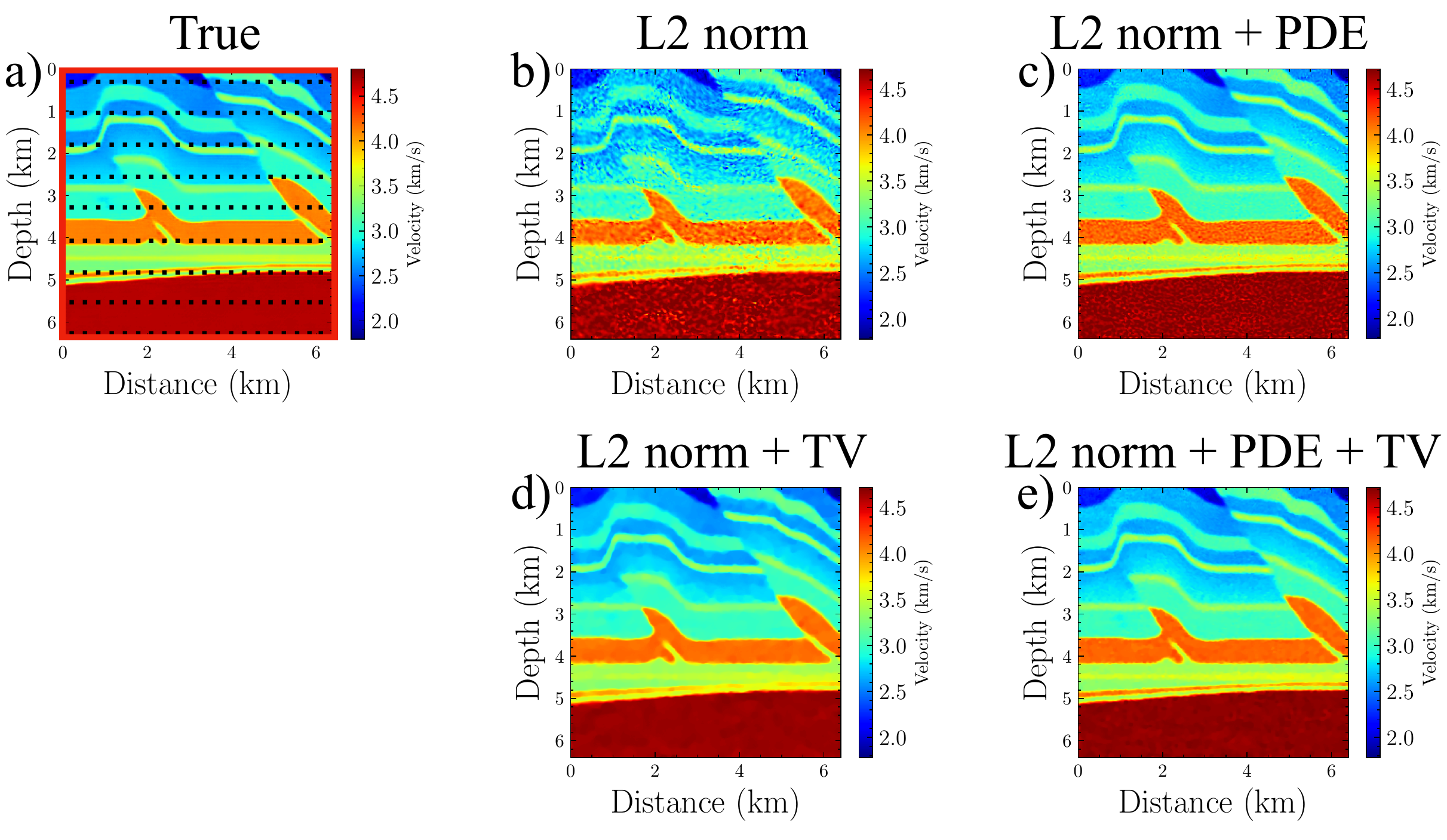}
    \caption{a) The true model. The inverted results with the observed wavefield recorded on the whole spatial domain using the (b) L2 norm objective function, (c) our proposed physics-informed FWI (L2 norm + PDE), and their further enhancement with total variation (TV) regularization (d and e, respectively). The red box denotes the region covered with receivers.}
    \label{fig:seg_fwi}
\end{figure}

The true velocity model, as well as the corresponding wavefield (real and imaginary parts) are shown in Figure~\ref{fig:fwi_seg_init_true} (top row). The bottom row contains the initial velocity as well as the background wavefield used as input to the neural operator.
We use an Adam optimizer with a learning rate of 0.1, and we set $\lambda$ to 6.5e-6.
After 100 iterations of neural operator-based FWI, we observe that although both methods adeptly reconstruct the velocity model, our approach achieves superior reconstruction quality, as shown in Figure~\ref{fig:seg_fwi}. 
With the physics-constrained loss, the inverted result is cleaner and more accurate. 
With TV regularization, we observe that our method still has improvements compared to the conventional neural operator-based inversion. The error of our proposed method is 0.0123, lower than that of the conventional neural operator method, which is 0.0148. This demonstrates again that the physics-informed loss can reduce artifacts that exist in the vanilla implementation. 

\begin{figure}[!htb]
    \centering
    \includegraphics[width=1.0\columnwidth]{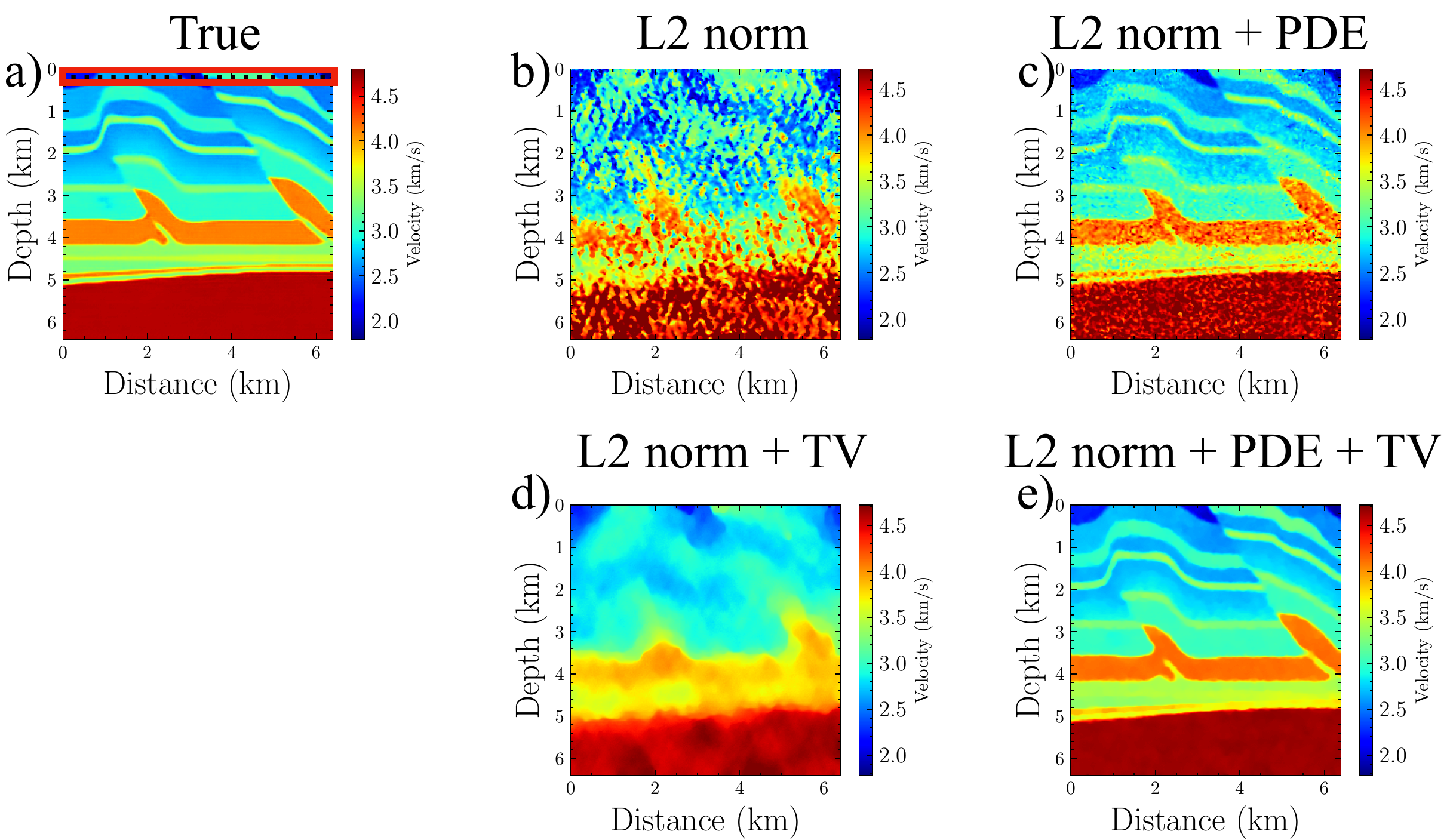}
    \caption{a) The true model. The inverted results with observations on the surface using the (b) L2 norm objective function, (c) our proposed physics-informed FWI (L2 norm + PDE), and their further enhancement with total variation (TV) regularization (d and e, respectively). The red box denotes the region covered with receivers.}
    \label{fig:fwi_seg_surface}
\end{figure}

Practically, we record the data on the surface.
Thus, we test the inverted results when observations are restricted to the surface, shown in Figure~\ref{fig:fwi_seg_surface}.
The experiments reveal that the conventional neural operator-based FWI, when applied to imperfect observations, struggles to accurately reconstruct the subsurface velocity model. Even with the application of Total Variation (TV) regularization, which improves the results by enhancing the low-wavenumber components, the method falls short of capturing the true subsurface structure.
In contrast, the physics-informed FWI approach not only recovers the main subsurface structure, but also provides a reasonably accurate estimation of the subsurface velocity. 
The addition of TV regularization to this method significantly improves and sharpens the inversion outcome, yielding results with cleanliness and precision.
These experiments underscore the efficacy of our proposed physics-informed FWI in handling both extensive and limited observational data.

\subsection{Tests on unseen velocity models with multiple sources and frequencies}
\label{unseen_curvea}
In the "Ideal tests" shown in Sections~\ref{ideal_curvea} and \ref{sec:ideal-ove}, where the neural operators' predictions are accurate enough, the physics-constrained loss can help to suppress the artifacts in the gradient, especially when the observation region is limited. 
However, due to the amount of training samples used for training and the model capacity, it is hard to reach the same accuracy level when applying the trained neural operator to the unseen velocity models.
As depicted in Section~\ref{ideal_curvea}, tiny prediction errors will introduce noise and artifacts. 
Those inherent errors in the wavefield prediction on the unseen target velocity models will introduce strong noise and artifacts in the inversion, resulting in failures of the inversion.
This is a challenge we always face when applying the neural operator-based FWI as we can not expect the inversion target to be seen during the training.
Hence, it is important to test the performance of the proposed method in such a scenario.

\begin{figure}[!htb]
    \centering
    \includegraphics[width=1.0\columnwidth]{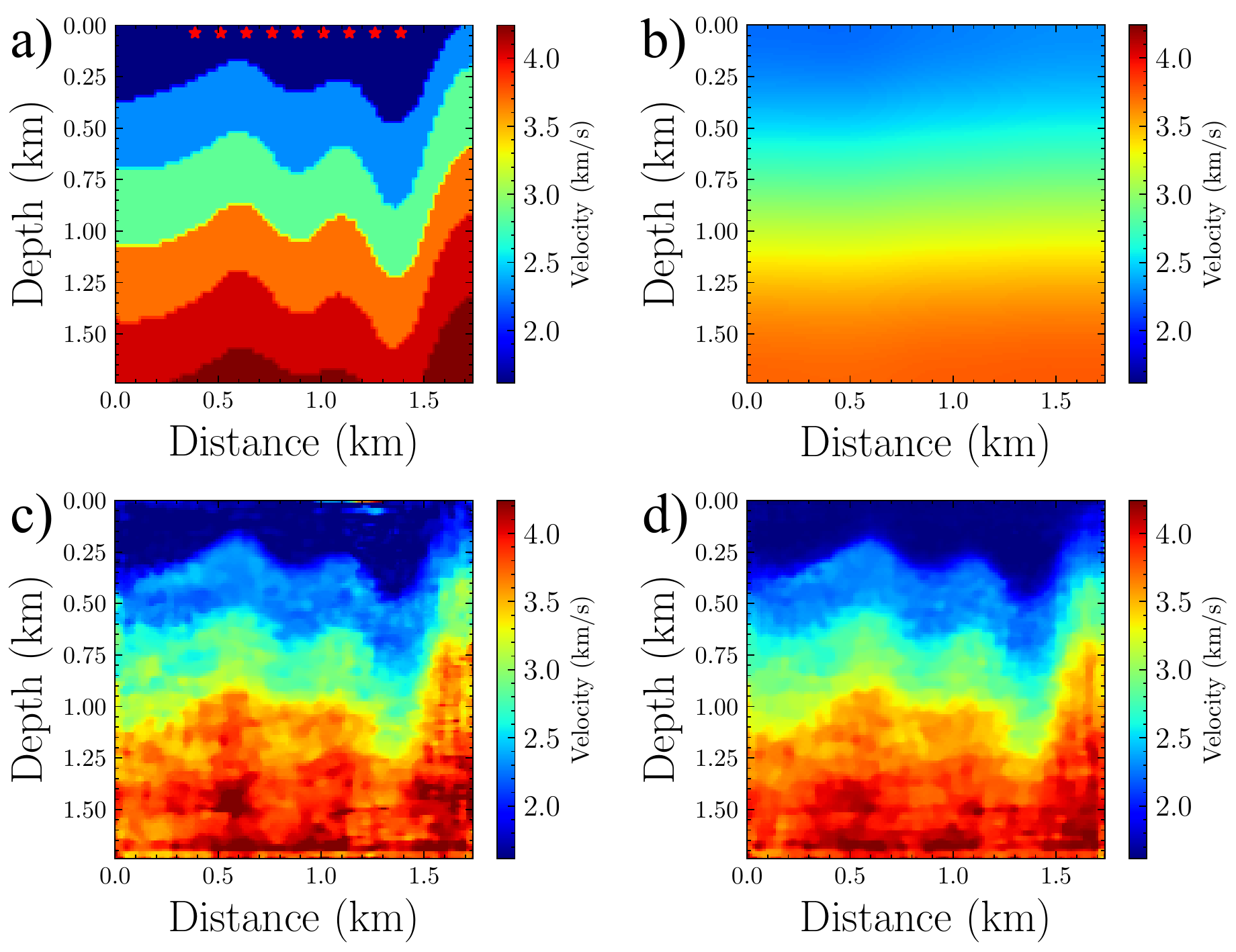}
    \caption{The true unseen velocity (a) and initial velocity (b), and the comparison of inverted results with TV regularization using the conventional neural operator-based FWI (c) and that using our proposed method (d).}
    \label{fig:fwi_multi_s_multi_freq_inv_comp_inv_369}
\end{figure}

We picked an unseen velocity model from "CurveVelA" velocity models, shown in Figure~\ref{fig:fwi_multi_s_multi_freq_inv_comp_inv_369}a. 
We tried the same setting of source signature (one shot and one frequency) in Section~\ref{ideal_curvea}; however, we found it is not stable to invert the velocity model even with observations on the whole spatial domain due to the inherent prediction errors of the neural operator on the target velocity model. 
Instead of focusing on a single frequency and single shot, we use multiple sources with frequencies ranging from 3 Hz to 12 Hz with a frequency interval of 0.5 Hz and using the observations from the whole subsurface domain.
The sources are uniformly located at a depth of 0.025 km, denoted by red dots in Figure~\ref{fig:fwi_multi_s_multi_freq_inv_comp_inv_369}a. 
We start from the initial model (Figure~\ref{fig:fwi_multi_s_multi_freq_inv_comp_inv_369}b) obtained by smoothing the true model. 
We use an Adam optimizer with a learning rate of 0.05 and set $\lambda$ to 1e-5, and compare the inverted results using conventional neural operator-based FWI (Figure~\ref{fig:fwi_multi_s_multi_freq_inv_comp_inv_369}c) and that using our proposed method (Figure~\ref{fig:fwi_multi_s_multi_freq_inv_comp_inv_369}d). For both approaches, we add TV regularization.
Although the improvements are not obvious as we saw in the "ideal tests", we still observe the surpassed artifacts in the inversion result from our method.
We also estimate the relative L2 norm errors of the inversion results compared to the ground truth to quantify the accuracy of the inversion results. 
The error of the inversion using our method is 0.0618, while that using the conventional approach is 0.0680. 
This demonstrates the improvement of the physics-informed loss term in the FWI.

\begin{figure}[!htb]
    \centering
    \includegraphics[width=1.0\columnwidth]{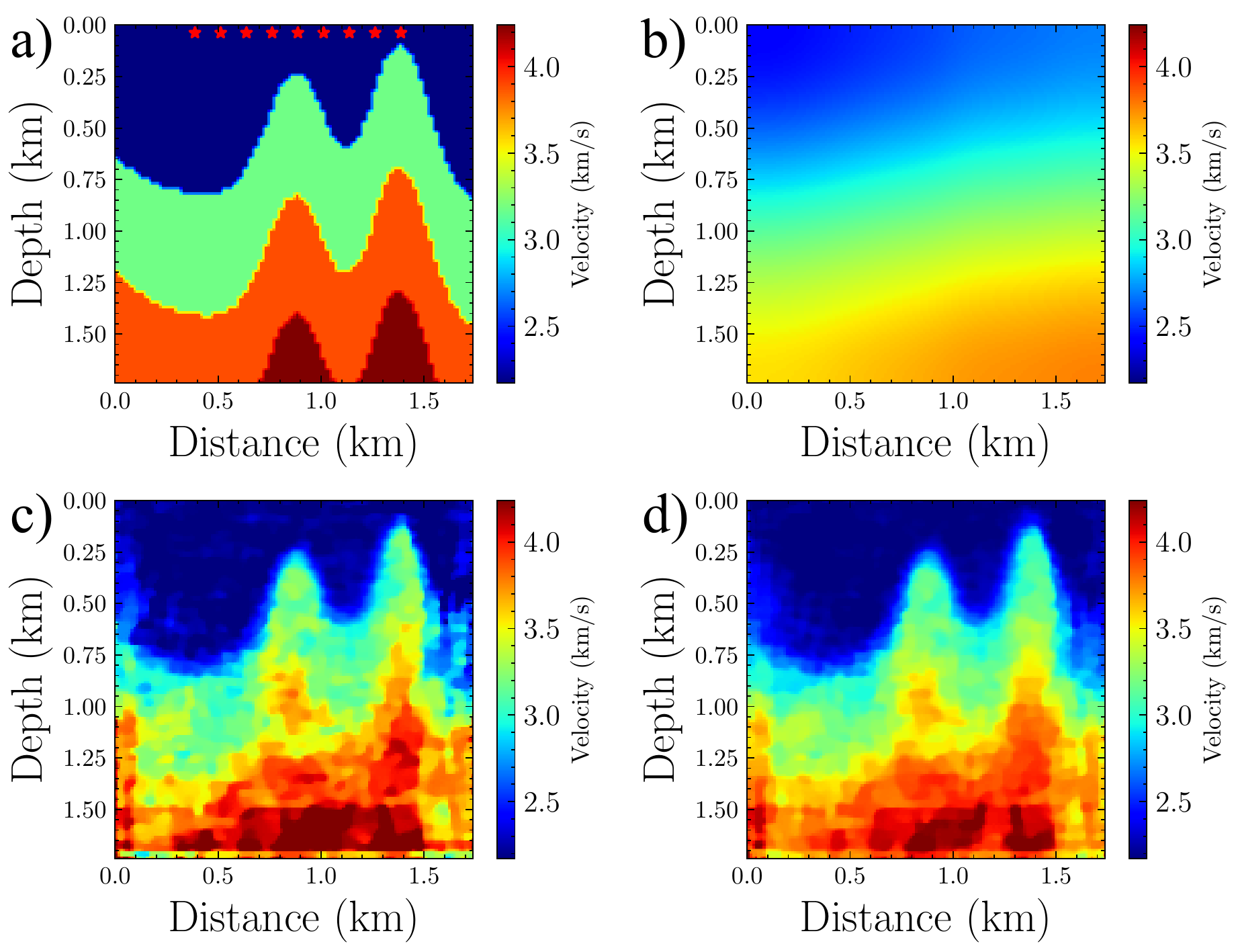}
    \caption{Another unseen case: the true velocity (a) and initial velocity (b), and the comparison of inverted results with TV regularization using the conventional neural operator-based FWI (c) and that using our proposed method (d).}
    \label{fig:fwi_multi_s_multi_freq_inv_comp_inv_429}
\end{figure}

We applied the same setting on another unseen velocity model, and the results are shown in Figure~\ref{fig:fwi_multi_s_multi_freq_inv_comp_inv_429}. 
Using our proposed approach, the artifacts and noise are somehow less than the conventional neural operator based FWI, especially at the deep part and the right-hand side part. 
The inversion result of our method shows a more reasonable reconstruction of the subsurface thanks to the incorporation of a physics-informed loss term. 
The error of the inversion result by our method is 0.0633, while that by the conventional approach is 0.0702. 
Those two examples demonstrate the improvement of our method in the scenario where the prediction error on the target unseen velocity is large.

\begin{figure}[!htb]
    \centering
    \includegraphics[width=1.0\columnwidth]{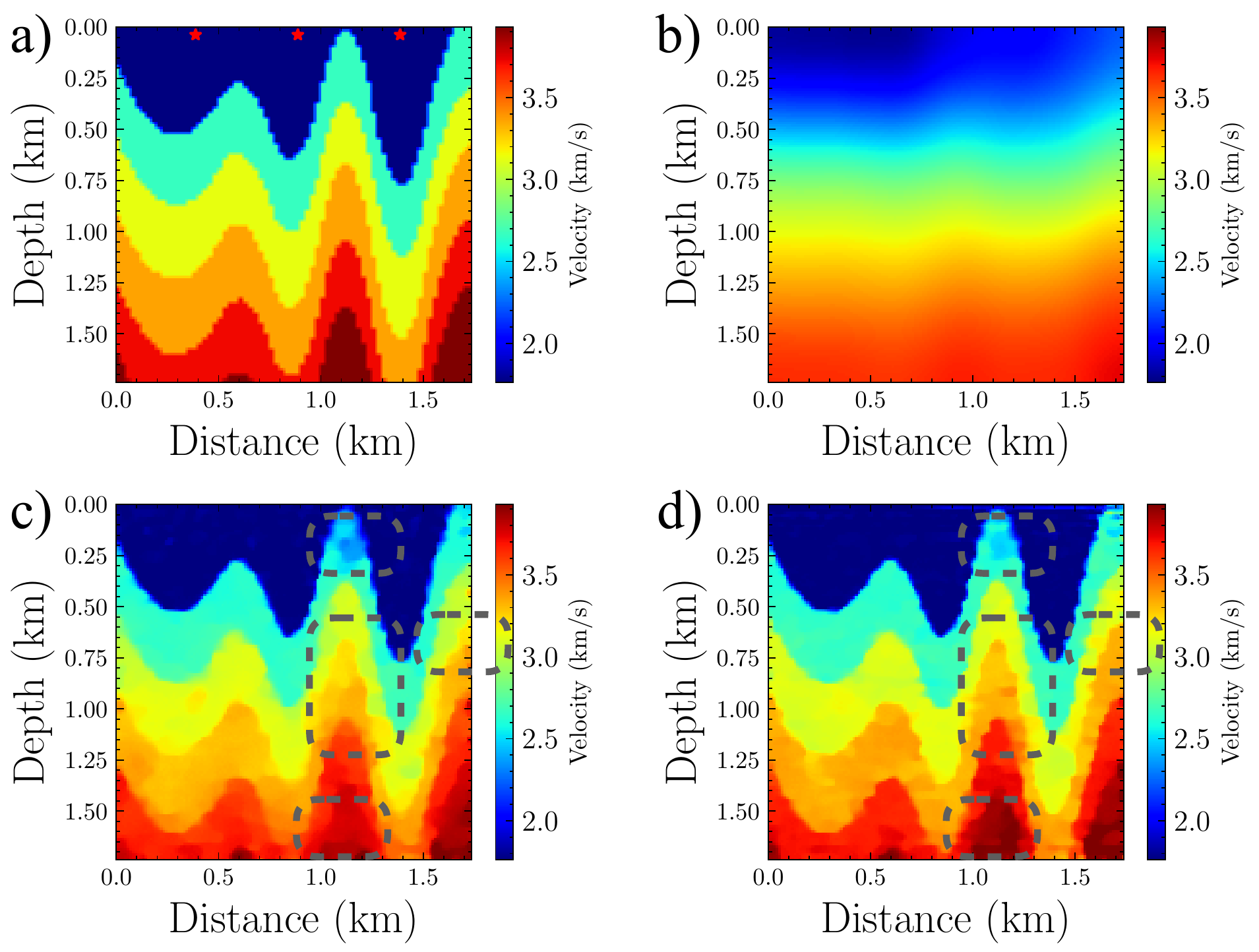}
    \caption{The true velocity (a) and initial velocity (b), and the comparison of inverted results with TV regularization using the conventional neural operator-based FWI (c) and that using our proposed method (d). The gray boxes denoted the obvious differences between the inverted results using different methods.}
    \label{fig:fwi_multi_s_multi_freq_inv_comp_inv_crime}
\end{figure}

In realistic applications, we expect to always face discrepancies between the observations and the predictions of the neural operator on the target velocity model. 
We also expect discrepancies for our numerically simulated data due to the physical assumptions we use, like acoustic or isotropic. However, we have additional limitations in neural operators, which are the approximate nature of the wavefield solution, and its dependency on the training set. If the training set is close to the model we expect the inverted model in real applications to look like, actually the neural operator can inject this prior information to help the inversion. On the other hand, if the training data is far away, like any prior, it can be detrimental to the inversion. 
In other words, by improving the accuracy of the neural operator-based prediction on the target velocity model and making it close to the observations, the improvements achieved by our method will be more obvious. 
To validate this, we utilize the "inversion crime" to test the proposed method.
We picked an unseen velocity model and placed three shots on the surface, shown in Figure~\ref{fig:fwi_multi_s_multi_freq_inv_comp_inv_crime}a.
We, then, predicted the surface data using the trained neural operator (used in Section~\ref{ideal_curvea}) as the observations, the frequency ranges from 3 Hz to 12 Hz with a frequency interval of 0.5 Hz. 
This is the scenario where the velocity model is not seen during the training, and the trained neural operator generalizes very well on the target velocity model even though it still does not generalize well on the intermediate velocity models during the inversion process.
Here, since the prediction errors exist in our neural operator, the physics, evaluated from the prediction on the target velocity model and the target velocity itself, is not consistent with the real physics, yielding large physics-informed residuals calculated on this prediction and target velocity.
Hence, optimization based on directly minimizing PDE residuals will not lead to the optimal solution.
To mitigate this inherent physics inconsistency, we incorporate the implementation of differential physics-informed loss. 
We simply use the physics-informed loss calculated using the current velocity model and wavefield predictions to subtract the reference physics-informed loss, which is obtained by the initial velocity model and observed data, to obtain the differential physics-informed loss.
To realize the calculation of the reference physics-informed loss, we set the measurements to be fully sampled along three rows of grid points at the top of this 2D velocity model.
We use an Adam optimizer with a learning rate of 0.05 and set $\lambda$ to 6e-6.
The inverted results are shown in Figures~\ref{fig:fwi_multi_s_multi_freq_inv_comp_inv_crime}c and~\ref{fig:fwi_multi_s_multi_freq_inv_comp_inv_crime}d.
Similar to our ideal tests, the improvements are more significant, and we marked the obvious ones using gray boxes. 
The error of the inversion using our method is 0.0213, and that of the conventional approach is 0.0282.
This scenario demonstrates that if the prediction on the target velocity model is accurate enough, even though the velocity model is not seen during the training, our method still shows obvious improvement.
In addition, it indicates that there is a possibility to improve the inversion accuracy by enhancing the generalization and accuracy of the neural operator-based forward modeling in cases where we did not have the possibility to see all velocity models. 

\section{Discussions}
\label{discuss}
Neural operator-based FWI significantly enhances computational efficiency compared to conventional numerical-based FWI, as neural operator-based simulations offer substantial efficiency and scalability advantages \cite{yang_rapid_2023, huang_learned_2024}. Although this approach requires an initial offline training phase, e.g., the “CurveVelA” training example required 15.75 hours (which can be reduced with a lightweight network), the subsequent inversion process benefits from instantaneous simulation, enabling highly efficient inversion workflows. 
Specifically, for the case illustrated in Figure~\ref{fig:fwi_multi_s_multi_freq_inv_comp_inv_crime}, where frequencies range from 3 to 12 Hz with a 0.5 Hz interval, the neural operator-based FWI takes only 30 seconds to complete 100 iterations. 
The proposed physics-informed FWI takes approximately 33 seconds for the same number of iterations, demonstrating that the computational overhead introduced by the physics-constrained loss and its corresponding backpropagation is negligible.
We shared the results of our proposed physics-informed FWI using neural operator-based wavefield solutions. 
These results demonstrated the potential of a physics-informed loss-based regularization to mitigate artifacts often encountered in neural operator-based FWI. 
Notably, our approach maintains robust performance even in scenarios where the vanilla neural operator-based inversion fails. 
In this section, we share insights into and limitations of the approach.

\subsection{Accuracy and generalization}
In theory, the primary difference between the proposed method and conventional FWI using numerical simulation is the substitution of the simulation engine with a neural operator.  
Consequently, the performance and optimization process of the proposed method should empirically approximate that of conventional FWI, and be affected by the accuracy of our prediction.  
Specifically, if the neural operator can accurately predict the wavefield for any given velocity, including perturbed versions of the target velocity, the proposed method could closely resemble numerical methods.
Currently, this level of accuracy is unattainable due to generalization issues, which presents a significant challenge for surrogate modeling. As a result, the performance of neural operator-based FWI will differ from conventional FWI, particularly during the intermediate stages of the inversion process, and the updates are not equivalent to those obtained using the adjoint state method. 
In this context, if we consider the neural operator as a conditional neural field, the FWI process can be viewed as the estimation of conditions (velocity models) given the observed data.

To further discuss this point, we would like to raise another question: {\it Do we really need the generalization to any velocity during the inversion process?}
While the neural operator will inevitably predict wavefields for velocities that lie between an initial model extracted from the ground truth and the ground truth and a foundational neural operator capable of handling all velocity models would be ideal to avoid this limitation of generalization, our primary goal in the inversion process is to obtain a final, reasonably accurate, result. This implies that the intermediate results are not as crucial, and there is no standard intermediate process. If we can ensure the accuracy of the target velocity models, the neural operator-based FWI will converge to the expected minimum. 

\subsection{The importance of the physics-constrained loss as an auxiliary tool to the generalization issue}
As previously discussed, the generalization of the neural operator to any unseen velocity model is not as effective as expected, and its predictions may sometimes violate wave physics. 
The tests in Section~\ref{unseen_curvea}, where the neural operator did not generalize well (with very low errors in the simulation) on any of the unseen (in training) velocity models, demonstrated that by adding physics constraints, the generalization of the neural operator-based FWI somewhat improves. 
However, as with most machine learning algorithms, the generalization will depend on the richness and solution representation of the training set, and in the case where the data are sufficient, the physics constraint might not be needed as it is well represented by the training data.
However, real-world cases often involve imperfect generalizations. Therefore, physics constraints serve as an essential auxiliary tool to address the generalization issue in surrogate modeling for intermediate velocity models, making them important for neural operator-based FWI.

\subsection{Utilize the priors in the neural operator by physics-constrained loss}
\begin{figure*}[!htb]
    \centering
    \includegraphics[width=0.9\linewidth]{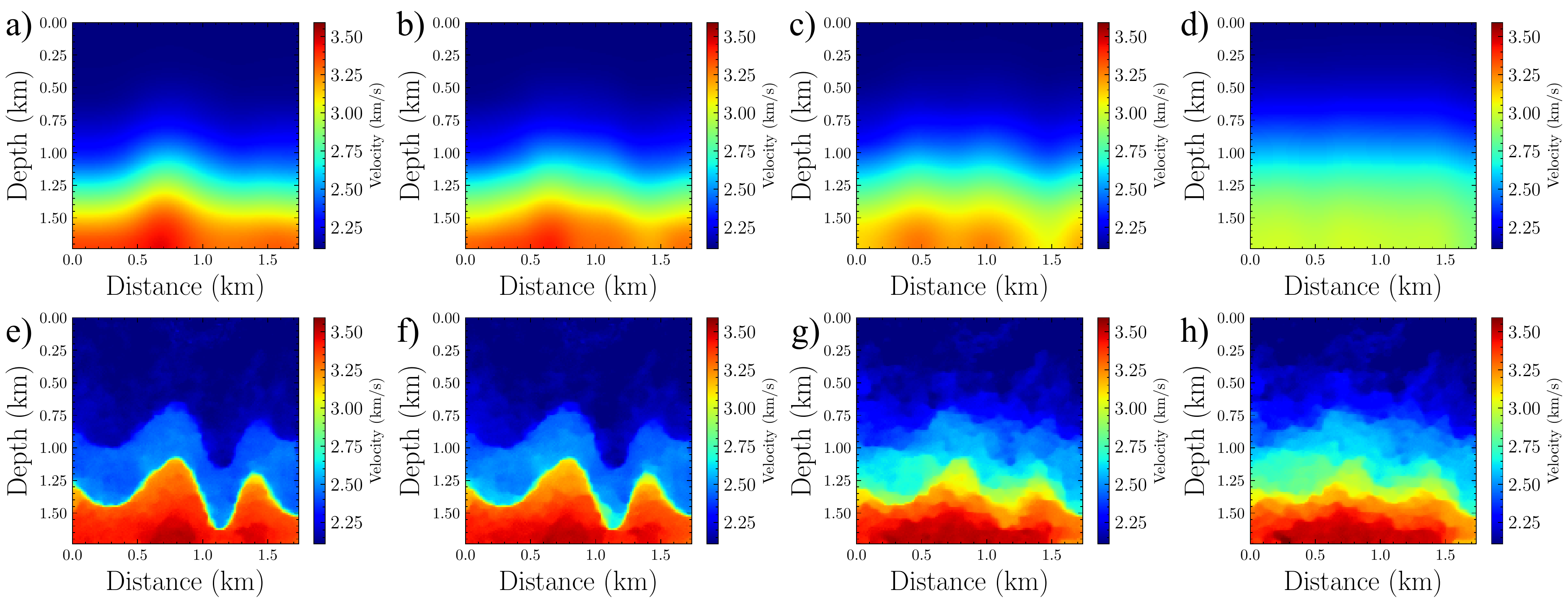}
    \caption{The inverted results (e-h) using PDE loss term and TV regularization only start from different initial models (a-d). The relative errors of the inverted result compared to the ground truth model for (e-h) are 0.021, 0.027, 0.059, and 0.073, respectively.}
    \label{fig:pde_loss_only}
\end{figure*}

Most surrogate modeling for the wavefield simulation can not be generalized to any velocity model. 
That means it is inherently biased by the distribution of the training velocity models.
We found that for this level of neural operators, adding the physics-constrained loss function in the FWI can recover the subsurface model with only the initial model, even without data constraints.
As shown in Figures~\ref{fig:pde_loss_only}e and f, starting from good initial models (Figure~\ref{fig:pde_loss_only}a and b), which is a smoothed version of the ground truth, only using the physics-constrained loss as the loss function can guide the neural operator-based FWI to recover high-quality subsurface models. 
It shows that the proposed method can utilize both a good initial model, which captures the main low-wavenumber information of the subsurface, and the priors stored in the neural operator, which is the biases on some specific velocity models through the training, to invert the velocity models even without the surface observed data.
However, what we want to highlight is not the fact that the proposed method recovered the subsurface model without data constraints. 
We want to highlight that the proposed method, specifically the physics-constrained loss in the misfit function, can fully utilize the priors of velocities stored in the neural operator implicitly.
As observed in Figures~\ref{fig:pde_loss_only}g and h, when the initial model is poor, the differences between the physics-constrained only FWI and FWI with both loss terms (Figure~\ref{fig:fwi_curvea_l2_pde_t v_inv_vs_l2_tv}) will be obvious.
Besides the influence of the initial model, another thing that should be noted here is that the prediction on the target velocity model should be accurate enough. 
Otherwise, the calculated PDE residuals using the target velocity model and the predicted wavefield will be large resulting in a failure of inversion with only a PDE loss, because the optimization of the PDE residuals to zero will not bring the initial model directly converge to the target velocity model. 
It can only compensate for the gradients and suppress artifacts in the inversion to some extent, as shown in Section~\ref{unseen_curvea}.
However, understanding how accurate the neural operator-based forward modeling should be remains challenging.

\begin{figure}[!htb]
    \centering
    \includegraphics[width=1.0\linewidth]{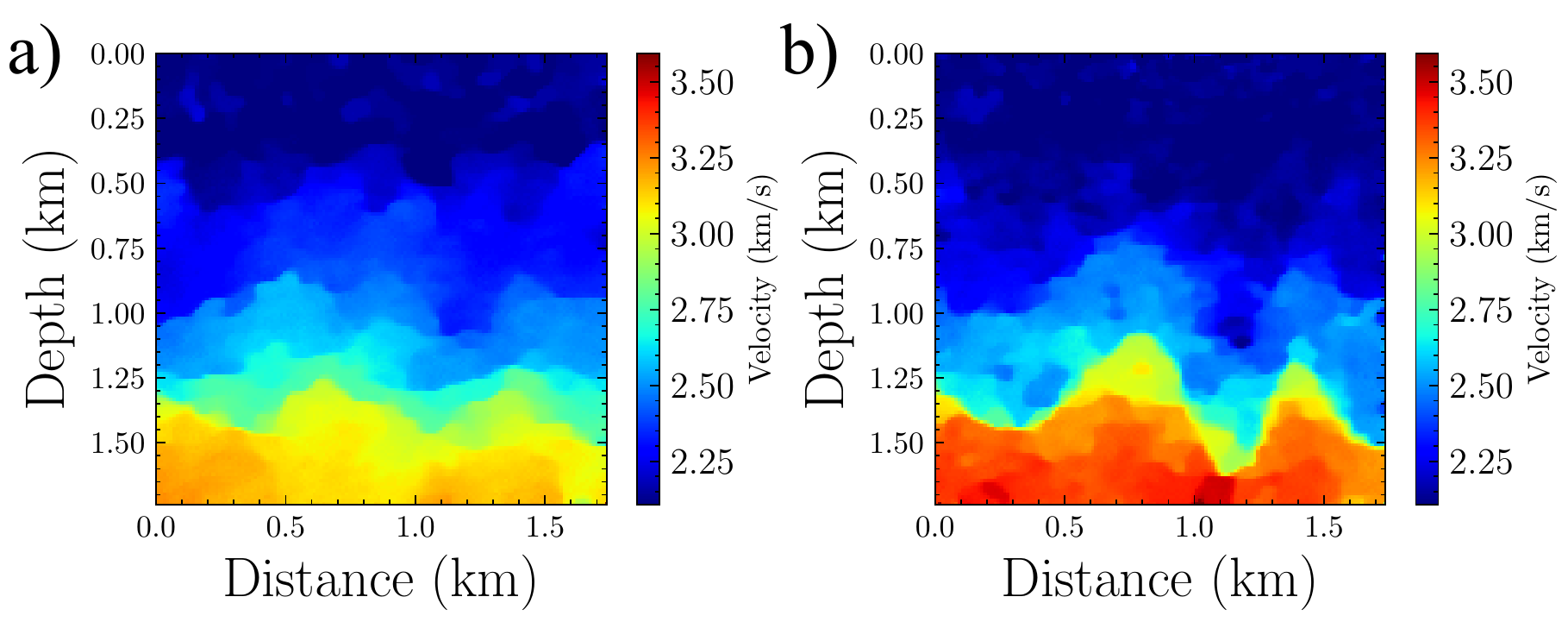}
    \caption{The inverted results of the conventional neural operator-based FWI (a) and that using our proposed method with PDE loss and data loss start from a poor initial model (Figure~\ref{fig:pde_loss_only}d).}
    \label{fig:comparison_on_very_smooth_model}
\end{figure}

In addition, to avoid the misunderstanding that the improvement of the inversion with limited (practical) observations (surface observations only) using our approach is completely coming from the prior extracted by the PDE loss term, in which the data do not play an important role in our proposed method, we pick Figure~\ref{fig:pde_loss_only}d as the initial model, and do the inversion with the observation sampled along thirty rows of points at the top of this velocity model.
The results are shown in Figure~\ref{fig:comparison_on_very_smooth_model}. 
In spite of the poor initial model, the FWI with only PDE loss and TV regularization can not reconstruct the subsurface velocity model. 
However, when combined with the observations, it generally recovers the subsurface velocity model. In contrast, the FWI with observation only fails to invert the subsurface velocity model. We argue that for neural operator-based FWI, combining the physics-constrained loss and data constraints would be the optimal choice.

\subsection{Further potential enhancements to the proposed method}
Despite the advancements in our proposed method, our findings also acknowledge room for further refinement to improve the generalization of the approach.
As stated above, the artifacts observed in the conventional neural operator-based inversion can be attributed to two primary sources: the inaccuracies inherent in the learned neural operator itself (especially for those unseen in the training models) and the errors introduced through the process of automatic differentiation due to the generalization ability of the surrogate modeling. 
Although those artifacts can be addressed to some extent with the addition of the physics-constrained loss term, we still observe noise in the gradient and inverted results using our method.
The noise can be removed by other conventional filtering means or regularization, and we have demonstrated further enhancement by means of TV regularization.
Beyond the TV regularization, we argue that there are additional potential enhancements like filtering the gradient or updating the velocity model in the latent space by dictionary learning \cite{Huang2019} or modern latent space machine learning techniques. 
Utilizing those techniques would also improve the inverted results and might help the inversion when the prediction errors on the target velocity model are large.

In addition, when the pretrained neural operator is applied to larger velocity models with resolutions different from those in the training dataset, the accuracy of wavefield prediction is lower than that obtained with the same resolution as the training data, resulting in degraded performance \cite{huang_learned_2024}. This issue can be partially mitigated by using the reference frequency strategy \cite{huang2022single} during the inversion process or by incorporating multi-resolution data into the training process of neural operators \cite{beyer2023flexivit}. 
The enhancement of the ability of our proposed method to handle unseen and even out-of-distribution models can come from the strategies mentioned above, and most importantly, is to improve the generalization capability of the neural operators themselves.

Another factor that needs to be considered when applying the proposed inversion approach to field-scale cases, where the frequency and model size may vary from those demonstrated in this paper, is the computational efficiency and scalability.
Since the forward modeling within our framework relies on neural-operator-based simulation, the overall cost of inversion is directly constrained by the efficiency and scalability of the underlying neural operator.
To investigate this, we benchmarked the FLOPs and memory usage under different resolutions with a batch size of 8.
For the baseline case (256$\times$256), training requires 52.23 GFLOPs per sample with a peak memory of 11.28 GB, while inference requires 17.41 GFLOPs per sample with a peak memory of 6.94 GB.
Doubling the resolution to 512$\times$512 increases the cost to 189.24 GFLOPs and 18.01 GB for training, and 63.08 GFLOPs and 13.87 GB for inference.
A representative field-scale setting (500$\times$1000) further raises the requirements to 358.84 GFLOPs and 31.92 GB for training, and 119.61 GFLOPs and 22.20 GB for inference.
Moreover, to maintain accuracy at higher frequencies, a naive solution is to increase the number of Fourier modes (e.g., from 48 to 64), which raises the cost for the 256$\times$256 case to 57.86 GFLOPs and 20.03 GB for training, and 19.29 GFLOPs and 6.44 GB for inference.
These results highlight the need for further improvements, such as lightweight network architectures for more efficient training and inference, like Vision Transformers with better scaling laws \cite{beyer2023flexivit}, or multi-scale FNO variants \cite{you2024mscalefno} to address frequency bias for improved high-frequency simulation.

\subsection{Potential extension to large-scale and complex media}
The examples presented in this paper focus on 2D, acoustic, and isotropic media, whereas realistic scenarios typically involve 3D, elastic, and even anisotropic media. 
Nevertheless, the proposed approach primarily aims to enhance the accuracy and stability of neural operator-based FWI, without being strictly dependent on the specific medium type or the neural network architecture involved. 
Thus, extending neural operator-based simulation methods to 3D, large-scale, and complex media is orthogonal to our proposed improvements, and these aspects can be effectively integrated to address more realistic scenarios. 
Although Eq.~\ref{sec8:l2normnew} is formulated for the acoustic case, the physics-informed loss can be generalized by using instead the elastic or viscoelastic wave equation, where the displacement (or stress) fields are required to satisfy the corresponding governing equations. In this setting, the loss naturally constrains both compressional and shear wave components, as well as their interactions across interfaces.
Since the FNO architecture is capable of handling multi-component fields, this extension enables the modeling of shear mode conversion and multiparameter coupling, and can also incorporate attenuation effects when viscoelastic terms are included.
However, we emphasize that the primary challenge for such an extension would lie in preparing extensive and representative training datasets required for training such a neural operator-based simulation network, which is very costly.
On the other hand, given that the computational cost of frequency-domain seismic modeling grows significantly with increasing frequency, finer grid resolution, and more complex governing equations, granted that advanced computational techniques such as multifrontal solvers \cite{amestoy2024recent} might reduce the computational cost and memory, the neural operator-based simulation will still exhibit more pronounced efficiency advantages compared to traditional numerical simulations, thus resulting in a substantially faster FWI approach, which makes this topic worth exploring.

Another promising extension of our approach is to address the local minima challenge in realistic complex media by integrating the proposed method within a global optimization framework. This integration could enable the generation of a high-quality initial velocity model, which serves as a strong starting point for subsequent numerical-method-based inversion. While global optimization methods, such as genetic algorithm, particle swarm optimization, and Bayesian optimization, can achieve the inversion starting from a very poor initial guess, like a constant background velocity model, they are often prohibitively expensive. 
The high efficiency of the neural operator-based inversion framework has the potential to mitigate this cost, making such hybrid strategies more practical for large-scale applications.

\section{Conclusion}
\label{conclusion}
By integrating the learned wavefield solution facilitated by the Fourier Neural Operator (FNO) with the physics-regularized loss function for Full Waveform Inversion (FWI), we successfully achieved an efficient FWI, producing results that are not only more accurate but also less noisy than the conventional neural operator based FWI. 
Even when faced with limited observational data, our method demonstrates robust subsurface reconstruction capabilities. 
The experiments on the CurveVelA models and Overthrust model demonstrate the effectiveness and superiority of the proposed method.
By analyzing the factors affecting the neural operator based FWI, we found that the prediction accuracy on the target velocity model is crucial for a reliable inversion.
Although the proposed method can improve the inversion results to some extent when the prediction errors are not small, the application of neural operator-based FWI to realistic cases remains challenging. 
Beyond our proposed method, the key should be to improve the generalization and accuracy of the neural operator-based simulation. 
In addition, understanding how accurate it should be requires further exploration.

\section{Acknowledgements}
We thank KAUST and the DeepWave Consortium sponsors for their support and the SWAG group for the collaborative environment. This work utilized the resources of the Supercomputing Laboratory at KAUST, and we are grateful for that.

\bibliographystyle{IEEEtran}
\bibliography{opl}
\ifCLASSOPTIONcaptionsoff
  \newpage
\fi
\end{document}